\documentclass{emulateapj}

\slugcomment{To appear in ApJ}

\shorttitle{Spatial Distribution Analysis in Young Clusters}
\shortauthors{Gutermuth et al.}

\begin{document}

\title{The Initial Configuration of Young Stellar Clusters: A $K$--band Number 
Counts Analysis of the Surface Density of Stars}

\author{Robert A. Gutermuth\altaffilmark{1}, S. Thomas Megeath\altaffilmark{2}, Judith L. Pipher\altaffilmark{1}, Jonathan P. Williams\altaffilmark{3}, Lori E. Allen\altaffilmark{2}, Philip C. Myers\altaffilmark{2}, \& S. Nicholas Raines\altaffilmark{4}}

\altaffiltext{1}{Department of Physics and Astronomy, University of Rochester, Rochester, NY 14627 (rguter@astro.pas.rochester.edu)}
\altaffiltext{2}{Harvard-Smithsonian Center for Astrophysics, Mail Stop 42, 60 Garden Street, Cambridge, MA 02138}
\altaffiltext{3}{Institute for Astronomy, University of Hawaii, 2680 Woodlawn Drive, Honolulu, HI 96822}
\altaffiltext{4}{University of Florida, Department of Astronomy, 211 Bryant Space Sciences Building, Gainesville, FL 32611}

\begin{abstract}

We present an analysis of stellar distributions for the young stellar
clusters GGD~12-15, IRAS~20050+2720, and NGC~7129, which range in far-IR
luminosity from 227 to 5.68~$\times 10^{3}~L_{\odot}$ and are all still
associated with their natal molecular clouds.  The data used for this analysis
includes near-IR data obtained with FLAMINGOS on the MMT Telescope and newly
obtained wide-field 850~$\mu$m emission maps from SCUBA on the JCMT.
Cluster size and azimuthal asymmetry are measured via azimuthal and radial
averaging methods respectively.  To quantify the deviation of the distribution
of stars from circular symmetry, we define an azimuthal asymmetry parameter and
we investigate the statistical properties of this parameter through Monte Carlo
simulations.  The distribution of young stars is compared to the morphology of
the molecular gas using stellar surface density maps and the 850~$\mu$m maps.
We find that two of the clusters are not azimuthally symmetric and show a high
degree of structure.  The GGD~12-15 cluster is elongated, and is aligned with
newly detected filamentary structure at 850~$\mu$m.  IRAS~20050+2720 is
composed of a chain of three subclusters, in agreement with \citet {chen97}, 
although our results show that two of the subclusters appear to overlap.  
Significant 850~$\mu$m emission is detected toward two of the subclusters, but 
is not detected toward the central subcluster, suggesting that the dense gas 
may already be cleared there.  In contrast to these two highly embedded
subclusters, we find an anti-correlation of the stars and dust in NGC~7129,
indicating that much of the parental gas and dust has been dispersed.  The
NGC~7129 cluster exhibits a higher degree of azimuthal symmetry, a
lower stellar surface density, and a larger size than the other two
clusters, suggesting that the cluster may be dynamically expanding
following the recent dispersal of natal molecular gas.  
These analyses are further evidence that embedded, forming clusters are often
not spherically symmetric structures, but can be elongated and clumpy, and that
these morphologies may reflect the initial structure of the dense molecular
gas.  Furthermore, this work suggests that gas expulsion by stellar feedback
results in significant dynamical evolution within the first 3~Myr of
cluster evolution.
We estimate peak stellar volume densities and
discuss the impact of these densities on the evolution of circumstellar disks
and protostellar envelopes.

\end{abstract}


\section{Introduction}

It is now commonly accepted that most stars form in clusters of 100 or more
stars \citep{ll03,porr03,carp00}.  In addition, young OB stars nearly always 
appear in groups and clusters of lower mass stars \citep{tpn99,mege02}.  
Clearly, understanding the process of forming stars in clusters is of
considerable importance.  Although we have an increasingly detailed picture of
relatively isolated star formation in nearby dark clouds such as those in the
Taurus complex, our understanding of formation in clusters has been hindered by
greater observational and theoretical challenges.

Observationally, the challenges of studying embedded clusters are their 
distance, their spatial density, and their association with high column density
molecular clouds.  While the Taurus complex provides a number of examples of
isolated star formation at 160~pc, most clusters are at distances of 300~pc or
greater \citep{ll03,porr03}, and thus require greater angular resolution and
sensitivity to study. 
The stars in these regions may be closely spaced, again requiring high angular
resolution to resolve the individual stars.  Finally, many clusters are deeply
embedded, and are thus obscured to observation at visible wavelengths.  For
this reason, the importance of young stellar clusters was not realized until
infrared cameras with detector arrays became available \citep{lada92}.   With 
the latest generation of wide field cameras on 6-10 meter class telescopes, 
near-IR observations can now probe a large sample of young clusters with the
sensitivity to detect objects well below the hydrogen burning limit, the 
angular resolution to resolve high density groupings of stars, and the field of
view necessary to observe the distribution of stars over multi-parsec
distances.  Furthermore, the {\it Spitzer} Space Telescope is now providing
detailed images of young clusters in the mid-IR, allowing us for the first time
to identify young stars with disks and infalling envelopes efficiently in
clusters out to 1~kpc and beyond \citep[e.g.,][]{mege04,whit04}.

Theoretically, the problem of star formation in clusters is perhaps an even
greater challenge, as a realistic analysis must include a network of non-linear
processes, such as turbulent motions and shocks in the natal molecular cloud,
potential collisions and multi-body interactions between young stars and
protostars, and the destructive influence of radiation and winds from the
newly formed stars.  Numerical models of turbulent fragmentation have been
successful at simulating the formation of clusters (or at least clusters of 
protostellar gravitationally bound cores) and these simulations are beginning
to yield a wealth of theoretical predictions which can now be investigated 
observationally \citep[e.g.,][]{kb00,khm00,bbb03,bbv03,gamm03,li04}.
For example, numerical simulations by \citet{bbb03} and \citet{bbv03} of 
turbulent, initially spherically symmetric, Jeans unstable molecular clouds 
produce collapsing, non-uniform filamentary density structures which are
formed through the dissipation of kinetic energy by hydrodynamic shocks,
allowing gravity to overcome turbulent support and force collapse.  In
these simulations, the gas eventually fragments into small ($<200$~AU), 
self-gravitating, sub-virial clumps, which are interpreted as protostars.  The 
models produce multiple groups of protostars which eventually merge to form
large clusters.  
However, these simulations have obvious limitations:
there are numerical problems associated with following the collapse of gas from
parsec to 100~AU size scales, the models assume idealized initial conditions, 
and the models 
also do not take into account the feedback from the newly formed stars.  In
particular, the later stages of the evolution in these models may have
questionable relevance to real clusters given that most young stellar clusters
do not evolve into bound clusters, in part due to the dispersal of most of the
binding mass before it forms stars \citep{ll03}.

With increasingly sophisticated numerical models beginning to predict
the spatial
distribution and densities of young stars, it is an appropriate time to explore
the configuration of young stellar clusters through wide-field infrared and
submillimeter imaging, both to test current models and to guide the development
of future models.  Through the analysis of the stellar surface density in young 
($<3$~Myr) stellar clusters still associated with their natal molecular gas,
and through a comparison of the corresponding dust and gas density structure,
we can study the initial configuration of young clusters and their subsequent
dynamical evolution.  We are now engaged in a multi-wavelength study of the
structure of young stellar clusters, including ground-based near-IR data,
{\it Spitzer} mid-IR imaging, and ground-based submillimeter and
millimeter-wave imaging \citep{gute04,mege04}.  
While testing cluster formation models is a long-term goal of the project, this
paper is primarily focused on developing quantitative techniques for
analyzing the spatial structure of young stellar clusters and applying these
techniques to new observations of three clusters.   Subsequent papers will
apply these techniques to the full sample of clusters observed in our survey.
It is our hope that this work will
encourage modelers to provide more accessible observable quantities from
their simulations.

In this paper, we present an initial analysis of near-IR and submillimeter
imaging of three clusters using near-IR images from FLAMINGOS on the 6.5~meter
MMT telescope and 850~$\mu$m continuum maps from SCUBA on the JCMT.  We
restrict our near-IR analysis to an $H-K$ color-based extinction analysis and
the distribution of sources detected in the
$K$--band.  The $K$--band is the most sensitive to embedded young stars
\citep{mege94}, and thus gives the most unbiased view of the distribution of
point sources.   Although the membership of individual sources cannot be
established by the $K$--band data alone, statistical methods can be used to
infer the overall distribution of young sources.  This approach is particularly
appropriate for studying the densest parts of young clusters, where the density
of young stars dominates over the density of field stars in the line of sight.
In a future paper, we will use data from {\it Spitzer}'s Infrared Array Camera
to find the distribution of sources with infrared excess emission in all
three of the regions presented; this approach is particularly effective in
probing regions with a lower density of young stars than the background stellar
density.

The three cluster regions chosen for this analysis exhibit signs of ongoing 
star--formation such as active molecular outflows, H$_{2}$O masers, and
associated IRAS point sources with significant far-IR luminosities
ranging from 227~to~$5.68 \times 10^{3}~L_{\odot}$ 
(see Table~\ref{littable}).  A brief summary of the current knowledge of
these three regions follows.

\subsection{GGD~12-15\label{ggd1215}}
GGD~12-15 is a young cluster region associated with the Monoceros molecular 
cloud, located at a distance of 830~pc \citep{hr76}.  It was originally 
identified from several patches of optical reflection nebulosity 
with associated Herbig-Haro (HH) objects \citep{ggd78}.  There is a wide-spread
(1.13~pc) stellar cluster detectable at near-IR wavelengths 
with an estimated 134~objects \citep{hoda94,carp00} and a moderate velocity 
bipolar molecular outflow oriented northwest-southeast centered on an
H$_{2}$O maser and secondary peak in 800~$\mu$m emission \citep{lhd90}.
A compact H~II region, with flux consistent with excitation by a B0.5 
zero--age main sequence star, was detected in radio continuum emission 
\citep{rodr80}.  The primary peaks of 800~$\mu$m dust emission 
\citep{lhd90}, $^{13}$CO emission, and C$^{18}$O emission \citep{ridg03} are 
also approximately centered on the H~II region.  Furthermore, a cluster of 
point-like radio continuum sources has been detected in the vicinity of the 
H~II region \citep{grg00,grg02}.  Some of these objects have near-IR and mid-IR
spectral energy distributions consistent with embedded protostars \citep{pt03}.

\subsection{IRAS~20050+2720}
IRAS~20050+2720 is a relatively small but very active site of embedded star 
formation originally identified as an IRAS point source.  It is associated with 
the Cygnus Rift, located at a distance of 700~pc \citep{wilk89}.  Multipolar 
outflows have been found centered on the main IRAS source position, one of
particularly high velocity (+/-~70~km/s) \citep{bft95}.  \citet{chen97} 
performed a near-IR investigation of this object (completeness limit 
of $K =$~16.2), and reported an 
estimated 100 cluster objects arranged in three tightly packed subclusters 
which they labeled A, B, and C.  The most reddened of these is subcluster~A, 
located at the IRAS point source position, as well as the peak of $^{13}$CO and
C$^{18}$O emission \citep{ridg03}.  The area around this subcluster has 
significant near-IR nebulosity, and two embedded sources were detected in
$L$--band (3.5~$\mu$m) and narrow-band $M$ (4.8~$\mu$m) imaging that were not 
detected at shorter near-IR wavelengths \citep{chen97}.  \citet{chin01} 
performed a submillimeter and millimeter study and found a strong emission
source at the IRAS point source position, with a ridge extending to the south. 
They also detected a diffuse ridge to the northwest at 1.3 mm and three 
point--like millimeter sources outside the 2.3$^{\prime}$ field of view of
their 450~$\mu$m and 850~$\mu$m maps.  These are likely protostars or
pre-stellar cores associated with the cluster.  

\subsection{NGC~7129}

NGC~7129 is a large region of reflection nebulosity that is located 1~kpc
from the Sun \citep{raci68}.  It is illuminated by two Be stars and an
associated cluster of lower mass stars \citep{hoda94}.  This cluster is located
within a cavity west of a kidney-shaped molecular cloud which is sharply
defined in both CO \citep{ridg03} and submillimeter emission \citep{fms01}.  
A nebulous filament which traces the cavity wall is the dominant feature of the
region at both optical and near-IR wavelengths.  The activity of forming young
stellar objects has been detected in several places within the molecular cloud. 
Within the cavity wall is a third massive star, LkH$\alpha$~234, and its
embedded companion \citep{wein96,cabr97}.  A jet discovered at visible
wavelengths \citep{ray90} with near-IR and mid-IR counterparts \citep{cabr97}
is centered on these two objects.  The jet is blue-shifted and points southwest
into the cavity, suggesting that it may be the counterjet to the red-shifted
molecular outflow to the northeast.  Another deeply embedded intermediate mass
protostellar object, FIRS2, is located three arcminutes to the south of the
main cluster \citep{epc98}.  FIRS2 is located at the region's primary peak in
$^{13}$CO emission \citep{bech78} and is associated with a multipolar
molecular outflow, suggesting that it may have one or more young companions
\citep{fuen01,misk01}.  Many other Herbig-Haro objects have been detected
around the region which are associated with sites of outflow activity
\citep{es83,hl85}.  Several large-scale outflows have been observed in molecular
hydrogen emission \citep{eisl00}.  {\it Spitzer} 3~to~8~$\mu$m and 24~$\mu$m
imaging has revealed a significant extended cluster of embedded Class~I~and~II
young stellar objects to the northeast of LkH$\alpha$~234 \citep{muze04,gute04}.

\section{Observations\label{obs}}

Near-IR observations in the $J$~(1.25~$\mu$m), $H$~(1.65~$\mu$m), and 
$K_{s}$~(2.15~$\mu$m) wavebands of NGC~7129\footnote{The $J$ and $H$ band
observations of NGC~7129 were originally presented by the authors in
\citet{gute04}} and IRAS~20050+2720 were obtained by
the authors on June~15-16,~2001 using the FLAMINGOS\footnote{We wish to
acknowledge the superb instrumentation developed by Richard Elston, who helped
transform the study of star formation during his too short life.} instrument
\citep{elst98} 
on the 6.5~meter MMT Telescope\footnote{Observations reported here were 
obtained at the MMT Observatory, a joint facility of the Smithsonian
Institution and the University of Arizona.}.  FLAMINGOS employs a 
$2048 \times 2048$ pixel Hawaii II HgCdTe detector array.  
The platescale of FLAMINGOS on the MMT is $0^{\prime\prime}.167$ pixel$^{-1}$, 
and thus the field of view is $5^{\prime}.7 \times 5^{\prime}.7$ per image.  
An overlapping $2 \times 2 + 1$ center position raster 
was used to make a $9^{\prime}.5 \times 9^{\prime}.5$ mosaic.  Five dithered
mosaics were obtained in each band with 15 seconds exposure time per image, for
a total exposure time of 75~seconds per band at $J$, $H$, and $K_{s}$.  
Similar observations in the $J$, $H$, and $K$~(2.18~$\mu$m) (instead of
$K_{s}$) wavebands of GGD~12-15 were obtained on February~23,~2002, also with
FLAMINGOS on the MMT Telescope.  The same raster pattern was 
used, but the exposure time per image was 30 seconds.  We obtained eight
dithered mosaics in $J$--band for a total of 240 seconds of exposure time, and
four each in $H$ and $K$--band for a total of 120 seconds of exposure time in 
each band.  
Conditions were photometric for all observations, and stellar point spread
function sizes ranged  
from $1^{\prime\prime}$ to $1.1^{\prime\prime}$ full width at half maximum
(FWHM) in all bands.  
Only the $H$ and $K$ ($K_S$) data are presented here.

Basic reduction of the near-IR data was performed using custom IDL routines,
developed by the authors, which include modules for 
linearization, flat-field creation and application, background frame creation 
and subtraction, distortion measurement and correction, and mosaicking.  Point
source detection and synthetic aperture photometry of all point sources were
carried out using PhotVis version 1.09, an IDL GUI-based photometry
visualization tool developed by R.~A.~Gutermuth.  PhotVis utilizes DAOPHOT
modules ported to IDL as part of the IDL Astronomy User's Library
\citep{land93}.  By visual inspection, those detections that were 
identified as structured nebulosity were considered non-stellar and rejected.
Radii of the apertures and inner and
outer limits of the sky annuli were $1^{\prime\prime}$, $2^{\prime\prime}$, and
$3.2^{\prime\prime}$ respectively.  FLAMINGOS photometry was calibrated by 
minimizing residuals to corresponding 2MASS detections, using only those 
objects with $H-K_{s} < 0.6$~mag to minimize color differences in the 2MASS 
and FLAMINGOS filter sets.  
RMS scatter of the residuals between the two
datasets ranged from 0.11 to 0.14 magnitude for the data presented.  Median
photometric uncertainty intrinsic to our FLAMINGOS measurements was 0.03
magnitude, and our adopted maximum uncertainty tolerance was 0.1 magnitude.  
Photometric magnitudes for stars which were brighter than the range of
magnitudes corrected by the linearization module were replaced with 2MASS
photometry.
The 90\% completeness limits are derived by adding successively dimmer sets
of synthetic stars created by using a Gaussian with a FWHM equal to that of the
observed point sources in each mosaic, extracting their fluxes using the
procedure described above, and rejecting objects with anomalous measurements.  
Measurements were considered anomalous if their photometric uncertainties,
$\sigma$, were greater than 0.1 magnitude, or if the recovered magnitude was
more than $3\sigma$ brighter than the input magnitude, indicating that the
synthetic star was coincident with a brighter star in the image.
The magnitude at which 90\% of
the stars are recovered is recorded in Table~\ref{complete}.  

The three clusters were observed with the SCUBA sub-millimeter camera at
850~\micron\ on the James Clerk Maxwell Telescope\footnote{The James Clerk
Maxwell Telescope is operated by The Joint Astronomy Centre on behalf of the
Particle Physics and Astronomy Research Council of the United Kingdom, the
Netherlands Organisation for Scientific Research, and the National Research
Council of Canada.} (JCMT) on Mauna Kea, Hawaii during a series of runs from 
June to November 2003 in moderate weather conditions
($\tau(850~\micron)=0.2-0.5$). The data were obtained in scan-mapping mode
whereby a map is built up by sweeping the telescope across the source while
chopping the secondary at a series of throws, $30^{\prime\prime}$,
$44^{\prime\prime}$, and $68^{\prime\prime}$. To minimize striping artifacts,
each set of three sweeps were performed alternately in right ascension and
declination. The SURF reduction package \citep{jlh98} was used to flat-field
the data, remove bad pixels, and make images. Although effective at removing
the high sky and instrument background, this observing technique necessarily
loses all features on scales larger than the largest chopper throw, 
$68^{\prime\prime}$. The calibration was performed by using the same observing
technique to observe Uranus (when available) or standard sources CRL~618,
CRL~2688, or IRAS~16293.  These data are presented here for morphological
comparison to stellar spatial distributions.  A full analysis of the SCUBA maps
will be presented in a future paper.

\section{Methodologies for Stellar Surface Density Analysis\label{methodology}}

Discriminating embedded young stars in clusters from field stars is difficult
since the colors of the young stars are similar to those of field dwarfs and
giants, and since reddening due to extinction dominates the colors of these
sources.  Young stars with strong IR-excesses from disks and envelopes can be
identified, and we will adopt this approach by combining {\it Spitzer} and
ground--based near-IR data in a forthcoming publication.  However, this
technique cannot identify diskless young stars.  An alternative is to use
statistical methods based on stellar surface
densities to isolate regions on the sky that are dominated
by cluster stars and measure several properties of the cluster stellar
distribution.  In the following subsections, we describe methods for isolating
cluster-dominated regions of sky, modeling field star contamination there, and
measuring several properties of the underlying stellar distributions of the
clusters.  We apply these methods to our observations of the three clusters.
In Section~\ref{akmaps}, we map the extinction through the molecular clouds
associated with each cluster, a key step in modeling the field star
contamination.  In Section~\ref{kmh}, we use $K$-band magnitude histograms
and the extinction maps to estimate the field star contamination and the number
of cluster members in the high density central regions of each cluster.
Azimuthally averaged stellar surface density profile fitting is used to
characterize the cluster sizes and locations in Section~\ref{radial}.  The
azimuthally averaged profiles prove to be incomplete characterizations of the
spatial distributions.  As shown by an analysis of the azimuthal distributions
of stars in Section~\ref{az}, there are significant deviations from circular
symmetry in two of the three clusters.
Finally, in
Section~\ref{gather}, we present a method for mapping stellar surface density
distributions using nearest neighbor distances, yielding another visualization
of the often asymmetric structure present in the stellar distributions of these
clusters.  

\subsection{Extinction Mapping\label{akmaps}}
Foreground and background stars are contaminants to studies of star clusters, 
obscuring the stellar distributions of interest.  Young cluster environments
are rich with dust, reducing the density of background stellar contamination.
However, the dust in these environments is often distributed non-uniformly,
making it far more challenging to accurately characterize the distribution of
detectable background stars.  

In order to correct for field star contamination, we must first map the
distribution of
extinction in our cluster regions.  The SCUBA 850~$\mu$m maps presented later
in this paper cannot be used alone for this, as dust emission which varies on
scales larger than the largest chopper throw ($68^{\prime\prime}$) is lost due
to the chopping sky subtraction (see Section~\ref{obs}).  Available CO emission
line maps \citep{ridg03} are also not ideal for this purpose for many reasons,
including low signal to noise in diffuse regions (C$^{18}$O), optical depth
issues leading to saturation in dense cores ($^{13}$CO), low spatial resolution
($50^{\prime\prime}$ beam size in both maps), CO depletion due to chemistry or
freeze--out in dense cores (both), or varying excitation temperature (both).
Instead, we derive the $A_K$ maps for this work using the $H-K$ colors of 
{\it all stars} detected in both bandpasses \citep{lada94,la01}.  Similar to
the stellar surface density mapping method described later on, we choose to
consider the $H-K$ values of the nearest $N=20$ stars at each point on a
uniform $2^{\prime\prime}$ grid to match the SCUBA map sampling.  
Using an interative outlier rejection algorithm, we compute the mean $H-K$
value from the $N$ nearest stars to that grid point.  The algorithm calculates
the mean and standard deviation of the $H-K$ values, rejects any stars
with an $H-K$ value $3\sigma$ from the mean, and then reiterates these steps
until the mean converges.
Outlier rejection in this
application should primarily reject effectively unextinguished foreground
stars, because the heliocentric distances to the clusters studied are fairly
small.  The resulting extinction map is convolved with a $15^{\prime\prime}$
Gaussian kernel to match the beam size of the SCUBA data and then converted to
$A_K$ using the reddening law of \citet{cohe81} 
($A_K = 1.385 ((H-K)_{obs} - (H-K)_{int})$) 
and an assumed average intrinsic color $(H-K)_{int} = 0.2$.

\begin{figure}
\epsscale{.5}
\plotone{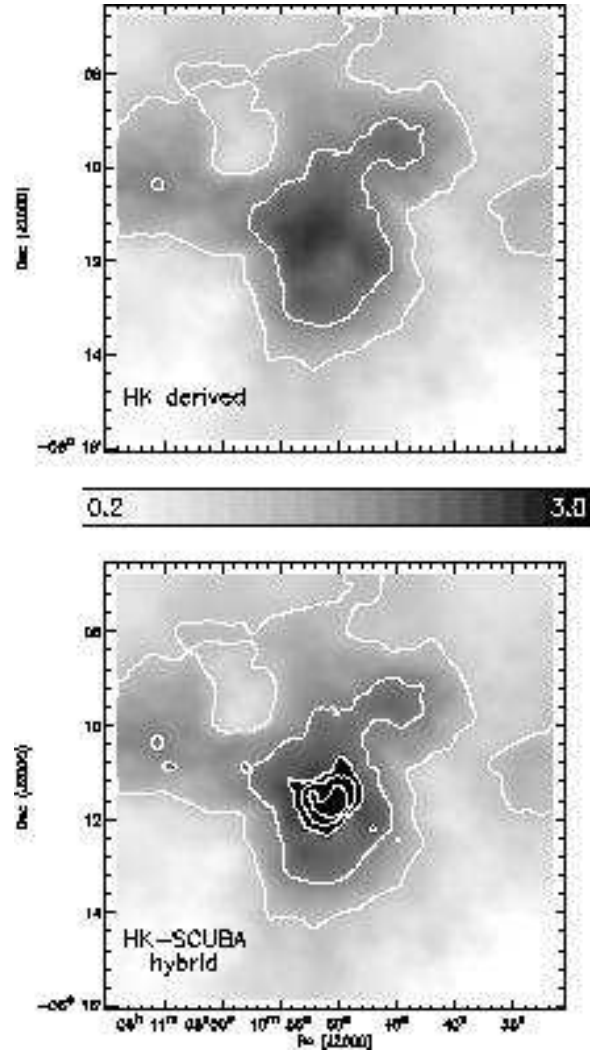}
\caption{Maps of $A_K$ as a function of position over the field of view of our 
near-IR observations of GGD~12-15.  The top map is derived purely from the
$H-K$ colors of all the stars detected in both bands.  The bottom map is the
same $HK$-derived map, updated with values derived from the SCUBA 850~$\mu$m
maps in places where there is a clear underestimate in the $HK$-derived values.
Note that these maps are log-scaled and inverted, with locations having
$A_K < 0.2$ in white and $A_K > 3.0$ in black.  The contour levels are 
$A_K = 0.6, 1.2, 2.4, 4.8, 9.6$.\label{ggd1215akmap}} 
\end{figure}

\begin{figure}
\epsscale{.5}
\plotone{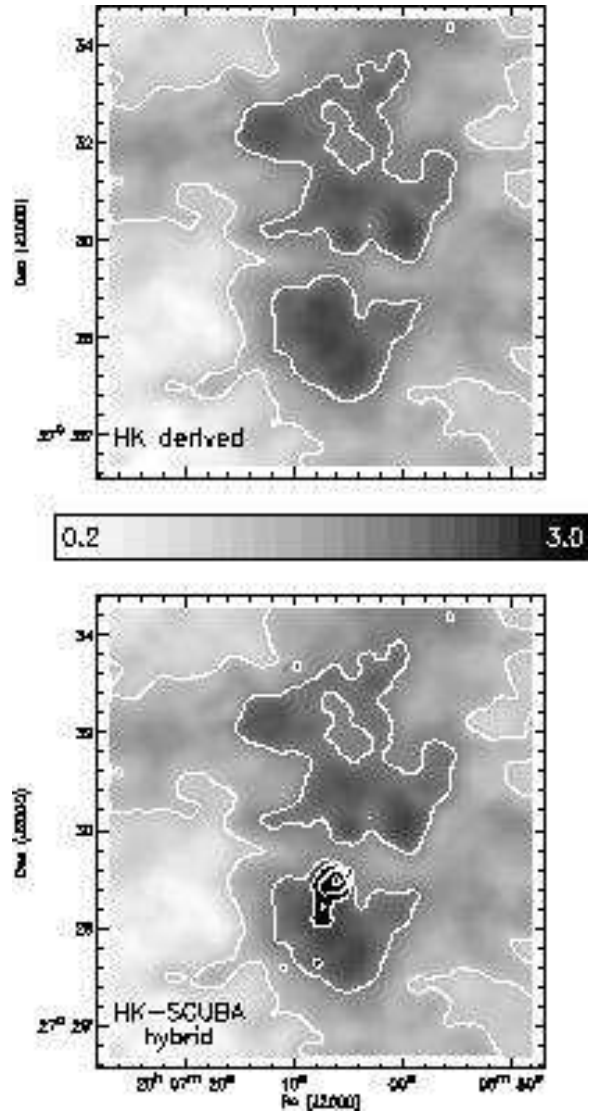}
\caption{Maps of $A_K$ as a function of position over the field of view of our 
near-IR observations of IRAS~20050+2720.  For a detailed description, see
Fig.~\ref{ggd1215akmap}.\label{iras20050akmap}}
\end{figure}

\begin{figure}
\epsscale{.5}
\plotone{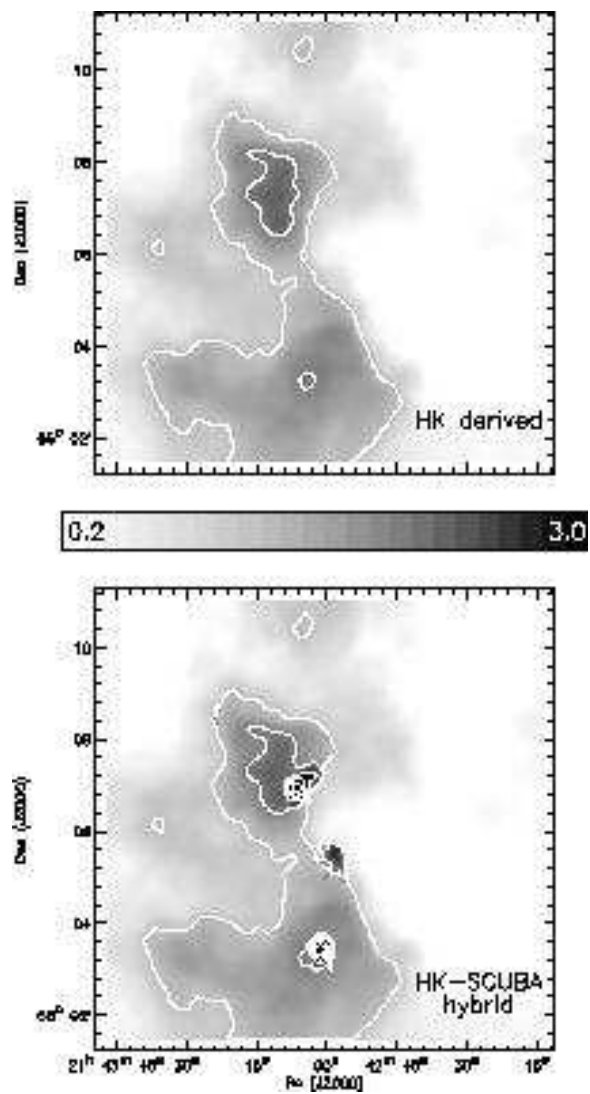}
\caption{Maps of $A_K$ as a function of position over the field of view of our 
near-IR observations of NGC~7129.  For a detailed description, see
Fig.~\ref{ggd1215akmap}.\label{ngc7129akmap}}
\end{figure}

This technique can underestimate the extinction since the average ($H-K$) color
can include a contribution from young stars embedded in the molecular cloud;
these stars have extinctions below the total line of sight extinction through
the cloud and bias the extinction toward lower values.  This bias is
particularly important in regions of high extinction, where the number of
detected background stars is small.
(Note that excess $K$--band emission from YSOs
will not have a significant effect compared to scatter about the assumed
intrinsic $H-K$ color.)  To address this bias, the SCUBA maps are used to
compare against the $A_K$ map and to improve it where necessary.  The SCUBA
fluxes are multiplied by $A_K/F_{850} = 0.272$~mag~pixel~mJy$^{-1}$
($2^{\prime\prime} \times 2^{\prime\prime}$ pixels) to convert
to $A_K$ and resampled onto the same grid as the $HK$-derived $A_K$ map.  The
$A_K/F_{850}$ conversion factor is derived assuming
$N_{H_2}/A_V = 10^{21}$~cm$^{-2}$~mag$^{-1}$ \citep[see Appendix D of][]{bm92},
$A_K/A_V = 0.090$ \citep{cohe81}, grain mass opacity
$\kappa_{850} = 0.01$~cm$^2$~g$^{-1}$ \citep{poll94}, and dust temperatures of
$T = 30 K$ for all three clusters \citep{lhd90,chin01,fms01}.  If any
SCUBA-derived $A_K$ measurements are more than 30\% above the
corresponding $HK$-derived $A_K$ values, the higher SCUBA-derived value is
used.  This hybrid $A_K$ map is used for 
the background contamination simulation.  The original $HK$-derived $A_K$ map
is preserved for use in a simulation of the clusters themselves, described in
Section~\ref{az} below.  Both maps are expanded and resampled using bilinear
interpolation to match the near-IR data in spatial scale and position on the
sky.  See Figs.~\ref{ggd1215akmap},~\ref{iras20050akmap},~\&~\ref{ngc7129akmap}
for the final extinction maps for GGD~12-15, IRAS~20050+2720, and NGC~7129,
respectively.

\subsection{$K$--band Apparent Magnitude Histograms\label{kmh}}

$K$--band Apparent Magnitude Histograms (KMH)\footnote{
To simplify the manuscript, all references to the FLAMINGOS $K$--band and 
$K_{s}$--band Apparent Magnitude Histograms will be referred to as
KMH regardless of the specific filter used.}  
are commonly used tools in the study of young stellar clusters.  
In the literature, these 
histograms are commonly referred to as $K$--band luminosity functions; for
clusters with a known distance and negligible intracluster extinction, each 
apparent magnitude bin corresponds to a unique luminosity range in the observed
wavelength band.  
The $K$--band is 
typically used since it is the least affected by extinction and the most 
sensitive to reddened young stellar objects.  Thus, the KMH has been used to 
study the initial mass function of young stellar clusters.  In this paper, 
we limit our analysis first to determining the magnitude at which the
contamination from field stars begins to dominate the $K$-band number counts,
and second, to estimating the total number of stars in each cluster.

We first define two regions in the observed field, the cluster-dominated region
of the field and an off-field, displaced from the cluster, where we can obtain
an estimate of the amount of field star contamination.  We assume that the
stars in the off-field are representative of the field stars over the entire
observed field.  The cluster-dominated region is defined via an analysis of
the radial density profile (see Section~\ref{radial}).  Since the determination
of the radial density profile also depends on a cutoff magnitude at which the 
field stars begin to dominate the number counts, several iterations were needed
to converge to a consistent magnitude cutoff and cluster region.
The off-field is selected based on relative lack of CO emission
\citep{ridg03}, lack of submillimeter emission, lack of significant or
structured extinction in the hybrid $A_K$ maps derived in Section~\ref{akmaps}
above, and overall lack of stellar surface density structure
(see Section~\ref{gather}).  Separate KMHs of the stars detected in both
regions are then constructed using a bin size of 0.5~magnitudes.  

To determine the magnitude at which the field star contamination begins to
dominate the number counts, we use the observed KMH of the off-field to model
the magnitude distribution of the stellar contamination in the cluster region.
We assume that there are no cluster members in the off-field.  To account for
the higher extinction toward the cluster, the background stars must be
artificially extinguished.  We separate the foreground and background
components of the off-field KMH by estimating the KMH of the foreground
stars using the \citet{wain93} galactic star counts model and subtracting
it from the observed off-field KMH.  We then measure the difference
between the mean cluster region and off-field extinction, $\delta A_K$,
using the hybrid $A_K$ maps (Section~\ref{akmaps}). This difference is
applied by shifting the background KMH by $\delta A_K$.  Then the
foreground contamination model is added to the shifted background model,
and the total is scaled by the ratio of the areas of the cluster region
and the off-field.  The observation--derived stellar contamination
estimates for the clusters presented agree to within statistical errors
with those derived from the galactic contamination model of
\citet{wain93}, once its background component has been shifted by the
average $A_K$ of the cluster region.

The resulting stellar contamination model histogram is subtracted from the
raw cluster KMH.  By inspection of the resulting difference histogram, we note
the magnitude at which the cluster region becomes dominated by field star
contamination.  Specifically, we use the faintest magnitude bin to have greater
than $3\sigma_{bkg}$ stars after field star subtraction, where $\sigma_{bkg}$
is derived by assuming that Poisson counting statistics holds for the number of
stars in each bin of the raw off-field KMH.  We adopt this magnitude as
the limiting magnitude for our data if it is brighter than the 90\%
completeness limit, otherwise the end of the first complete magnitude bin
brighter than the 90\% completeness limit is used.  Stars fainter
than the limiting magnitude are not included in the stellar surface density
analyses presented in this work. 

\begin{figure}
\epsscale{0.8}
\plotone{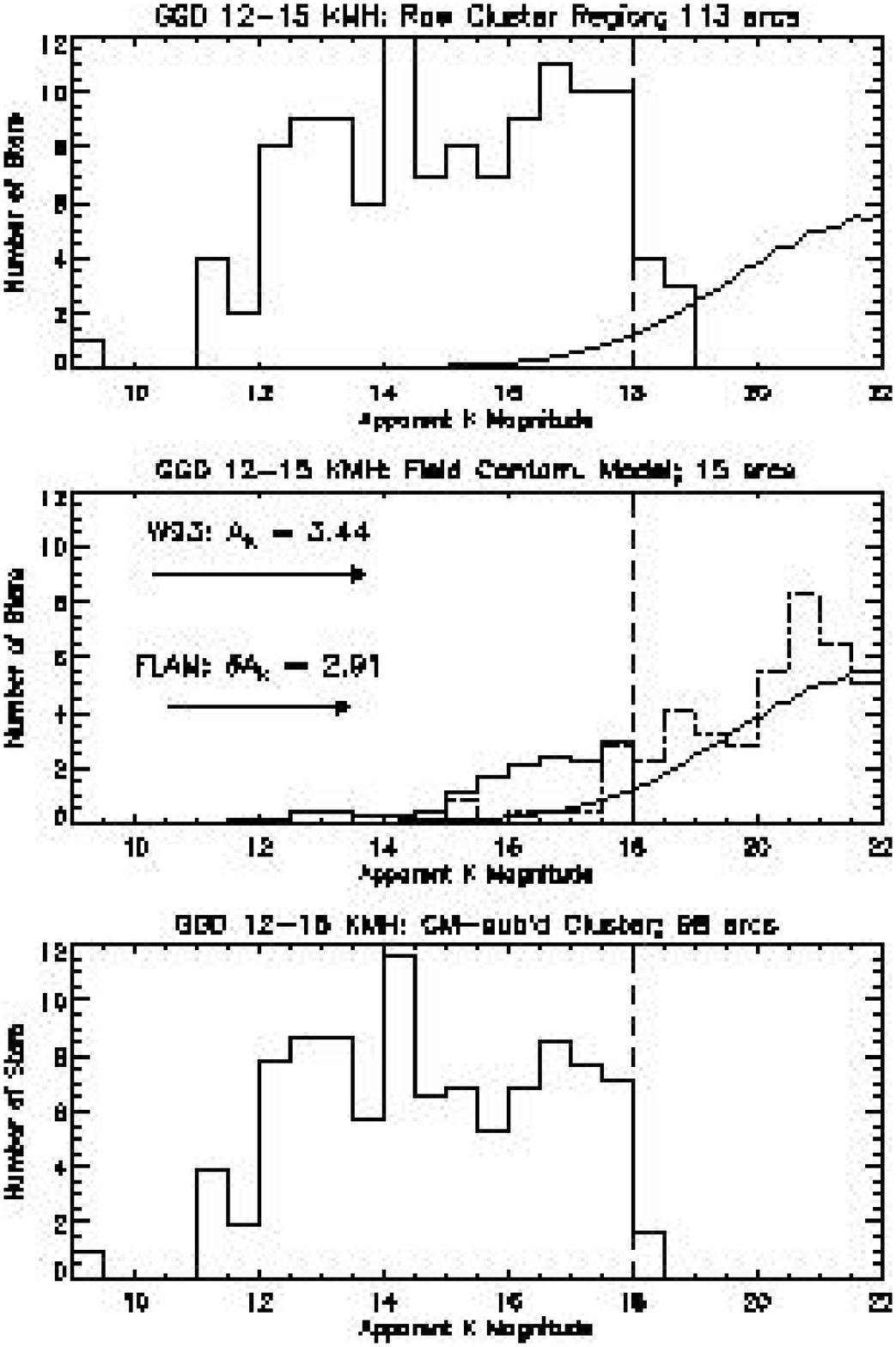}
\caption{$K$--band apparent magnitude histograms for GGD~12-15.  The top plot 
consists of all stars detected within the cluster region boundary.  The 
middle plot shows three field star KMH models.  The dot-dashed line is the
measured off-field KMH, shifted by $\delta A_{K}$ and scaled by the ratio of
the areas of the cluster core and off-fields.  The smooth, solid line
overplotted on this and the first plot is the field star model of 
\citet{wain93}, shifted by the average $A_{K}$ of the cluster region, for 
comparison.  The solid histogram is the mean KMH of the stars within the
cluster core boundary from the $N_{CM}=1000$ Monte Carlo field star models
generated in Section~\ref{kmh}.  The final plot is the difference of the raw
cluster core histogram and a hybrid of the two field star KMHs derived from our
observations.  The mean Monte Carlo model KMH is used up to the limiting
magnitude denoted by vertical dashed lines in all three plots.  Beyond this
limit, the scaled and shifted off-field KMH is used.\label{ggd1215klf}}
\end{figure}

For the purposes of determining the total number of stars and studying the
spatial distribution of stars in each cluster, a more sophisticated treatment
of the field star contamination is required which takes into account the
spatially varying extinction.   To estimate the position dependent magnitude
distribution of field stars, we construct Monte Carlo
simulations
of the foreground and background stellar distributions in each of the regions
observed.  We use the KMH measured from the off-field, minus the expected
foreground KMH from the model of \citet{wain93}, to determine the number of
background stars per magnitude per square pixel of the image plane.  These are
assumed to have mean extinction $A_{K,0}$, measured directly from the off-field
of the hybrid $A_K$ map, as before.  We then uniformly populate an image
plane equal in angular size to our observations with a Poisson-deviate
number of stars with mean equal to the total of the off-field background KMH,
appropriately scaled to the area of the image plane.  The magnitude of each
star is randomly sampled from the off-field KMH and the appropriate
positionally dependent extinction, 
$\delta A_K = A_K[x,y] - A_{K,0}$, is applied to each star.  All stars dimmer
than the adopted magnitude limit for the given cluster observations are
rejected.  To this sample we add a similar uniformly distributed population of
unextinguished foreground stars, using the \citet{wain93} model foreground KMH
to determine their density and magnitudes.  We perform $N_{CM}$ iterations of
the contamination model, where we have chosen to use $N_{CM}=1000$ for this
work.  These iterations are combined for use in the final field star
contamination characterizations which will be used throughout the rest of this
work.  Depending on the dust distribution, these simulations can predict a
higher amount of contamination than that obtained using an average $A_K$ (for
example, see the middle plot of Fig.~\ref{ggd1215klf})

The final contamination histogram is measured directly from the stars that fall
in the cluster region of our simulated field star distributions, and divided by
$N_{CM}$.  This is subtracted from the raw cluster KMH we observed.  The star
counts in Table~\ref{clustertable} are calculated by totaling the
contamination--subtracted cluster region histogram for all magnitudes brighter
than the limiting magnitude adopted (see the final plot in
Fig.~\ref{ggd1215klf}).  

\subsection{Radial Density Profiles\label{radial}}

Many studies of young stellar clusters in the literature have analyzed 
azimuthally averaged radial profiles of stellar surface density 
\citep[e.g.,][]{carp97}.  This method allows the resulting trend to be fit
by a wide variety of functions and naturally yields a size estimate for the
cluster.  Unfortunately, physical interpretations derived from these analyses
are often suspect because many young clusters appear to exhibit non-uniform and 
asymmetric structure
\citep{ll03}.  As we will show, azimuthal averaging of elongated 
clusters can lead to deceptively smooth, well behaved radial density profiles.  
Thus for this study, we limit our radial density profile analyses to estimation
of the size and location of the cluster core and cluster-dominated region.

To generate azimuthally averaged radial stellar surface density profiles for
each cluster, we first choose a center 
point that falls approximately at the center of the projected extent of the 
stellar cluster.  This choice is then refined by taking the mean of the 
Right Ascension and Declination of all stars detected within one arcminute of 
the initially chosen center point.  Based on this refined center point, we 
count the number of stars in concentric annuli of equal radial extent and 
divide by the area of 
each annulus.  The radial bin size used in this work is equivalent to 
0.034~pc at the assumed cluster distances.  It was chosen to achieve adequate 
spatial resolution while maintaining acceptable statistical weight for the 
inner stellar density measurements.  Given $N_i$ stars in the $i^{th}$ 
annulus, the stellar surface density for that radial bin is computed as 
follows:
\begin{displaymath}
\sigma_i = \frac{N_i}{\pi (r_i^2 - r_{i-1}^2)}
\end{displaymath}

We fit the measured radial surface density histogram, weighting each bin by
the number of stars it contains, using a function of the form:
\begin{displaymath}
\sigma(r) = a_0 + a_1 e^{(-r/r_0)}
\end{displaymath}

\begin{figure}
\epsscale{1}
\plotone{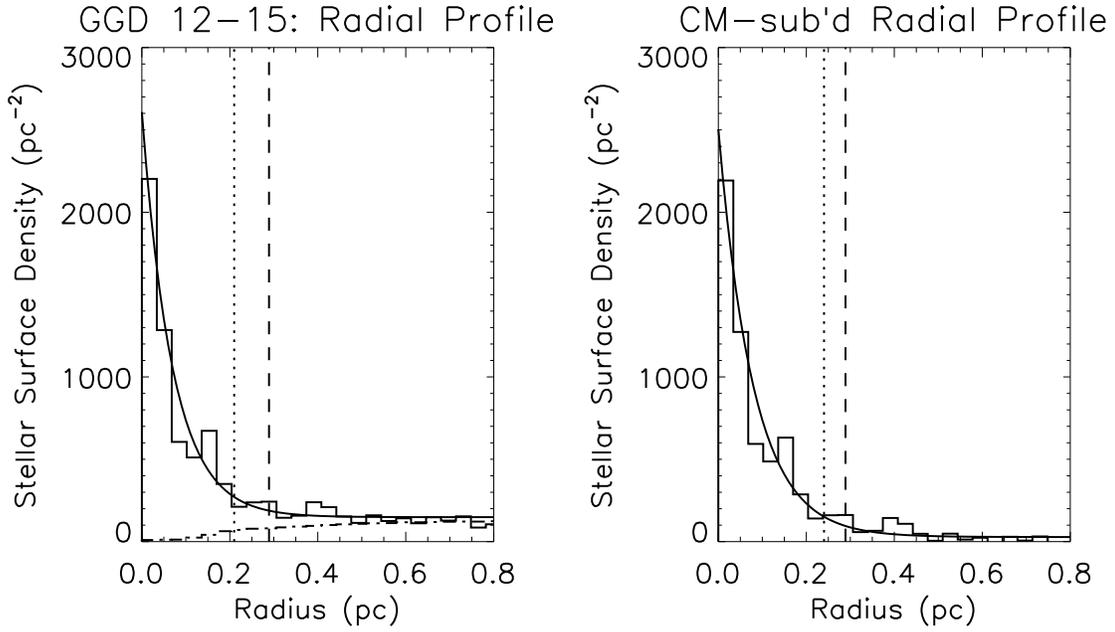}
\caption{Azimuthally averaged stellar radial density profiles for the GGD~12-15
region.  The solid line histogram on the left shows the raw measured radial
profile.  The overplotted dot-dashed histogram is the mean radial profile from
the $N_{CM}=1000$ Monte Carlo field star models.  The solid line histogram on
the right is the contamination--subtracted radial profile.  Overplotted on both
are smooth solid lines showing the exponential plus constant fits, vertical
dashed lines marking the radii where the source counts are $3\sigma$ above the
field star counts, and vertical dotted lines marking the cluster core radii, 
three times the e-folding lengths from the fits (see
Section~\ref{radial}).\label{ggd1215rad}}
\end{figure}

This function implies that field star contamination has a uniform
surface density $a_0$.  As mentioned in Section~\ref{akmaps}, background
stellar densities will vary significantly in regions of variable extinction.
Using the stellar distributions from the Monte Carlo contamination
models produced in Section~\ref{kmh}, we measure azimuthally averaged radial
profiles from the combined $N_{CM}$ stellar contamination models using the
identical bin sizes and center points used for the observations.  The resulting
densities are then divided by $N_{CM}$, and subtracted from the corresponding
observation-derived densities.  These contamination--subtracted profiles are
then fit with the above function, ensuring that the cluster profiles are
properly characterized independently of nonuniform background contamination 
(cf. Fig.~\ref{ggd1215rad}).
  
We define the cluster-dominated region as a circle centered at the refined
center point with radius of the point at which the contamination--subtracted
profile fit descends below $3\sigma$ above the average residual field star
density (see Table~\ref{clustertable}), where $\sigma$ is computed directly
from the outermost six radial bins. 
This isolates the region of elevated stellar
density, and thus the extent of the region that is dominated by cluster
members.  We note empirically that three times the e-folding length, $r_0$,
approximates this distance well in the two centrally condensed clusters
presented, but not in the case of a more diffuse cluster.  This value is also
noted in Table~\ref{clustertable} as the {\it cluster core radius}.  

\subsection{Azimuthal Distribution Histograms\label{az}}

Radial density profiles and azimuthally averaged distributions of the stellar
density in clusters do not take into account the 
asymmetric and often clumpy structure observed in many young stellar clusters 
\citep{ll03}.  Indeed, such methods actually erase asymmetric structure 
which may yield vital insight into the process of clustered star formation.  
To examine the azimuthal distribution of sources and determine whether the
clusters presented show a high degree of azimuthal asymmetry, we construct
azimuthal distribution histograms for each of the three clusters. 
Specifically, we count the number of stars in the cluster-dominated region that
are located in $45^{\circ}$ position angle bins sampled at $22.5^{\circ}$,
the Nyquist rate.  

To quantify the azimuthal asymmetry of a cluster, we first compute the
standard deviation of the histogram about the mean number of stars per
azimuthal bin.  We assume that this is the combined statistical noise of the
cluster distribution and the field star distribution.  We define the 
{\it azimuthal asymmetry parameter} ($AAP$) of a young stellar cluster as the 
ratio of the measured standard deviation of the azimuthal distribution to the
square root of the mean number of stars per azimuthal bin.  Given $m$ uniform 
overlapping $(360/(m/2))^\circ$ azimuthal bins with $n_i$ stars in the $i^{th}$
bin and mean $|n| = N/(m/2)$ stars per bin, the raw $AAP$ is computed as
follows:
\begin{displaymath}
AAP_{raw} = \frac{\sqrt{\sum_{i=1}^m{\frac{(n_i - |n|)^2}{m-1}}}}{\sqrt{|n|}}
\end{displaymath}

A circularly symmetric cluster has a random azimuthal distribution.  Therefore
its azimuthal distribution histogram should be uniform to within Poisson noise,
as should a uniform distribution of contaminating field stars.  This implies
that a combination of both also has a uniform azimuthal distribution.
Given successful isolation of a subset of stars in a
region which is dominated by cluster members, either via the statistical
methods described above or via true population restriction such as the
consideration of stars exhibiting excess infrared emission only, a quantitative
statistical argument for the presence of significant asymmetric structure can
be made using the measured  $AAP$.  Note that better isolation of cluster
members reduces additional noise from the random field star distribution,
yielding a stronger statistical argument.   

One additional complication is that variable extinction can introduce asymmetry
in the background stellar distribution.  This must be accounted for in order to
properly isolate any asymmetry that is due to the cluster distribution.
Similarly to Section~\ref{radial}, we can measure the distribution of stellar
contamination in azimuth using the Monte Carlo simulations performed in
Section~\ref{kmh}.  We measure the azimuthal distribution histogram of the
stars that fall in the cluster region boundary of the $N_{CM}$ contamination
models, and divide each bin by $N_{CM}$.  Subtracting this mean model
distribution from the observed one removes the systematic asymmetry expected in
the background stellar distribution, but lacks the appropriate Poisson scatter.
Thus it is important to note that noise from the field star distribution is not
removed by this method.  Given $b_i$ stars in the $i^{th}$ bin of the field
star model azimuthal distribution and mean field stars per bin of $|b|$, the
corrected $AAP$ is the measured standard deviation of this subtracted
distribution over the square root of the mean of the {\it unsubtracted}
distribution, $|n|$, plus a small contribution from the Poissan noise of the
mean field star model itself:
\begin{displaymath}
AAP_{sub} = \frac{\sqrt{\sum_{i=1}^m{\frac{(n_i - b_i - |n-b|)^2}{m-1}}}}{\sqrt{|n|+|b|/N_{CM}}}
\end{displaymath}

In order to argue for the presence of significant asymmetric structure, we must
detect azimuthal stellar surface density variations larger than that which is
likely via Poisson counting statistics alone.  To determine the probability
that a measured $AAP$ is consistent with a random azimuthal distribution, we
have performed Monte Carlo simulations of $10^6$ young cluster fields for each
of the regions presented.  The foreground and background components of each 
simulated field are generated identically to our field star contamination
models, as described in Section~\ref{kmh}.  In addition,
the clusters themselves are also modeled.  For the reference magnitude
distribution for the clusters, we have chosen to use the well-sampled 
$K$-band luminosity function (KLF) of the Orion Nebula Cluster (ONC) from 
\citet{muen02}.  The exponential fits of the subtracted radial profiles (see
Section~\ref{radial}) are used as the reference radial density distributions
for our model circularly symmetric clusters.  The number of cluster members in
a given model is a Poisson-deviate of the integrated exponential component of
the radial density profile fit.  The stars should be distributed uniformly in
position angle about the measured cluster center point of the
observations, but we must account for reduced effective sensitivity to low mass
cluster members in higher extinction areas.  To do this, we compute an
azimuthal distribution weighting function for a given radius by integrating the
assumed KLF up to the adopted magnitude limit minus the $HK$-derived $A_K$
value at that radius over all position angles.  The normalized results yield
the position angle probability distribution functions for the range of radii
possible in our field. 

Each of the $10^6$ young cluster fields generated is analyzed identically to
the observations.  The refined cluster center points are computed, using the
center point measured from the observations as the initially chosen point which
is then further refined.  Radial profiles are measured and the contamination
models used to treat the original observations are also used to compute and
remove radial variations in stellar contamination.  The resulting radial
profile is fit as described in Section~\ref{radial}, with those simulations
with nonconverging fits rejected as unphysical and regenerated from the
model\footnote{Nonconvergent profile fits most often occur in clusters with low
ratios of peak surface density to field surface density, such as NGC~7129 in
this work.  Since the resultant radial profile of a given iteration of the
model is allowed to vary according to Poisson scatter, clusters with less
central condensation may produce stellar distributions that are
indistiguishable from flat density plateaus or other non-exponential
geometries.  Model stellar distributions that take on such profiles which lack
central peaks and exponential character are unlikely to be identifiable as 
cohesive clusters distinguishable from field stars by our stellar surface
density isolation methods, and therefore we reject those iterations.}.
The cluster-dominated region is thus defined as in Section~\ref{radial}.  The
azimuthal distribution is then measured, and we use the same mean of $N_{CM}$
stellar contamination models, recentered on that field's refined cluster
center point, to remove any systematic variation due to structured
extinction.  The $AAP$ of each iteration is then computed, compiled into a
histogram, and then normalized to yield the probability distribution of the
$AAP$ over all the simulations.  

\begin{figure}
\epsscale{1}
\plotone{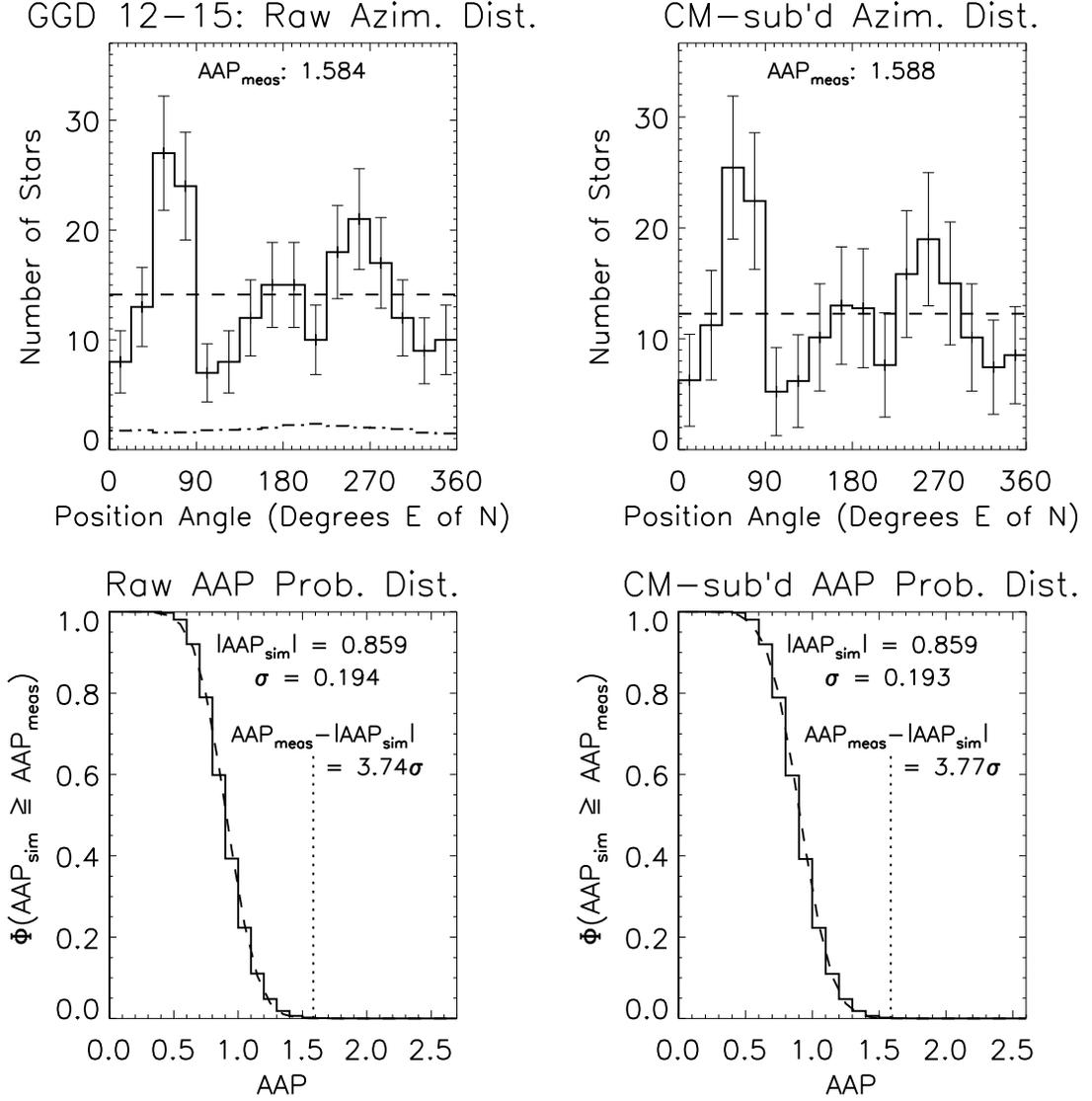}
\caption{{\bf TOP}: Azimuthal distribution histograms of all stars within the
cluster core boundary of GGD~12-15.  The horizontal dashed line marks the mean
number of stars per bin, corresponding to an ideally uniform stellar
distribution for reference.  The dot-dashed histogram plotted on the left is
the mean azimuthal distribution of the $N_{CM}=1000$ Monte Carlo field star
models.  
{\bf BOTTOM}: Probability distribution histograms for the $AAP$ derived from
$10^6$ model cluster fields (see Section~\ref{az}).  The left plot shows the
raw $AAP$ distribution, and the right shows the $AAP$ after subtraction of the
recentered mean azimuthal distribution of the field star models.  Overplotted
on both is a dashed line corresponding to a Gaussian of mean $|AAP_{sim}|$ and
standard deviation $\sigma$ to demonstrate that the distributions are
effectively approximated this way.  Vertical dotted lines mark the location of
the appropriate measured $AAP$.\label{ggd1215az}}
\end{figure}

From these distributions, we can determine the probability that the measured
$AAP$ of the cluster of stars observed is consistent with a random azimuthal
distribution, and is thus an approximately circularly symmetric cluster.  A
high measured $AAP$ suggests a very small probability that the cluster is
circularly symmetric, implying the presence of significant asymmetric
structure.  
As an example, inspection of the 
GGD~12-15 probability distribution in Fig.~\ref{ggd1215az} suggests that a
measured $AAP$ of 1.588 has a $\sim$0.008$\%$ probability of being consistent
with a circularly symmetric distribution.  Thus the presence of azimuthal 
asymmetry can be argued for clusters with measured $AAP$ values significantly 
greater than $|AAP_{sim}|$, while those with $AAP$ values close to 
$|AAP_{sim}|$ are consistent with a circularly symmetric distribution.  
The resulting field star model subtracted $|AAP_{sim}|$ values from the
simulations for GGD~12-15, IRAS~20050+2720, and NGC~7129 were $0.859$, $0.869$,
and $0.894$ respectively.

\subsection{Stellar Surface Density Maps\label{gather}}

Ultimately, a stellar density distribution mapping method must be employed to 
investigate the relationship between forming stars and their natal molecular 
gas distribution.  
Given the wide range of stellar surface densities in a young cluster region, a
method that employs adaptive smoothing lengths is ideal for achieving both high
dynamic range density measurements and high spatial resolution in locations of
high stellar surface density.  
For this analysis, we have chosen to use a variation on the nearest neighbor
method \citep{chri98} to construct the stellar surface density maps for each
cluster. 
This method exhibits many of the benefits of the GATHER algorithm outlined by 
\citet{glad99}, but lacks weighted averaging of multiple smoothing lengths.  
At each sample position $[i,j]$ in a uniform grid we measure $r_N(i,j)$, the
projected radial distance to the $N^{th}$ nearest star.  For convenience, we
chose a grid spacing of $3^{\prime\prime}$ for all stellar surface density maps
presented in this work.  The local stellar surface density at each grid
position $[i,j]$ is computed as follows:
\begin{displaymath}
\sigma(i,j) = \frac{N}{\pi r_N^2(i,j)}
\end{displaymath}

\begin{figure}
\epsscale{1}
\plotone{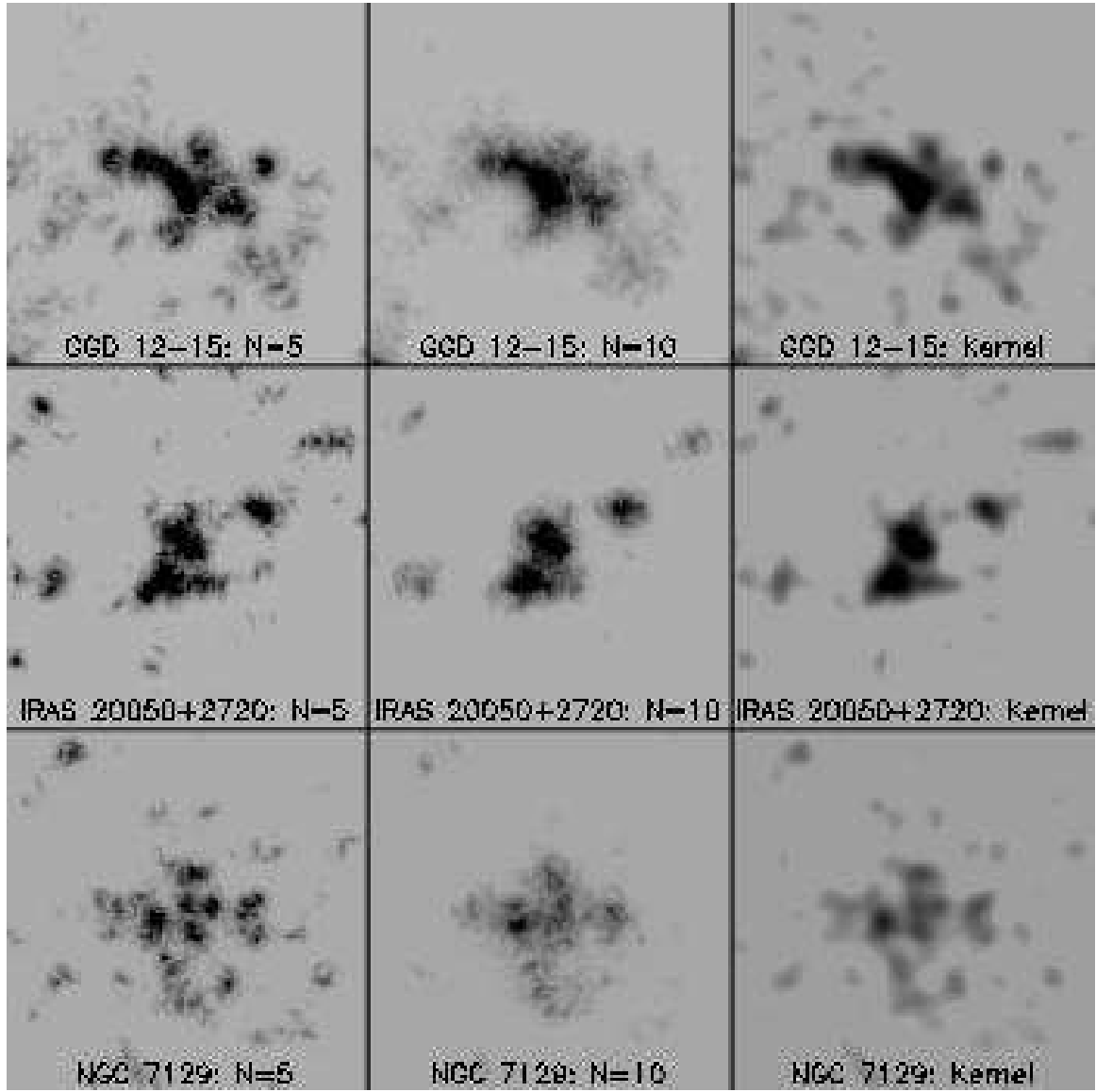}
\caption{Comparison of three different stellar surface density mapping methods 
for the three clusters presented: GGD~12-15, IRAS~20050+2720, and NGC~7129.  
In the left column are the $N = 5$ stellar density maps used throughout this 
work (see Section~\ref{gather}).  The maps in the central column are made with 
the same method, but the densities at each grid point are calculated using the
distance to the nearest $N = 10$ stars.  Maps derived using Gaussian kernel
smoothing \citep{gome93} with a smoothing width of $8^{\prime\prime}$ are in
the right column.  All maps are $5^{\prime}$ on a side.\label{dens3x3}}
\end{figure}

In Figure~\ref{dens3x3}, we present the inner $5^{\prime}$ of the stellar 
density maps derived with this method, using $N = 5$ and $N = 10$ (left and 
center columns respectively) to demonstrate how the maps change as a function 
of $N$.  The most obvious difference between the two is loss of spatial 
resolution with larger $N$.  A more subtle difference is the total loss of 
some small, high density features in the $N = 10$ maps.  Any high density 
subgroups that contain less than $N$ stars are often lost or significantly 
diminished with this method, thus it is critical to choose $N$ to be small
enough to allow small, high density subgroups to be properly represented in the
map.  Furthermore, $N$ must be large enough that the measured density
structure is not dominated by binary or triple systems or random
coincidences.  For
these reasons, we have chosen to use maps derived for $N = 5$ for this study.  
Peak surface density entries in Table~\ref{clustertable} were obtained directly
from the $N = 5$ maps.  For independent comparison, the right column of
Fig.~\ref{dens3x3} has equivalent grid resolution stellar density maps derived
using Gaussian kernel smoothing \citep{gome93} with a smoothing length of
$8^{\prime\prime}$.  
Note that the density distribution morphologies from the $N = 5$ maps are
discernable using this commonly used alternative method.

Approximate volume densities can be derived from surface density measurements
only by making assumptions about the distribution of stars in the line of
sight.  
Since our definition of a cluster--dominated
region is spatially confined to a circle, one might naively assume that the
stars were distributed evenly along the line of sight over a length of the
order of the cluster core radius.  
However, toward peaks in the observed surface density, it is unlikely that the
stars are distributed uniformly over the cluster diameter along the line of
sight, and this density would be a lower limit to the peak volume density.
An alternative method is to derive volume density
(see Table~\ref{clustertable}) using the nearest neighbor distance $r_N(i,j)$
defined above, thus assuming local spherical symmetry at those sample
positions.  We use $N=5$ for these measurements.  For grid position $[i,j]$,
the nearest neighbor volume density is computed as follows:
\begin{displaymath}
\rho(i,j) = \frac{N}{\frac{4}{3} \pi r_N^3(i,j)}
\end{displaymath}
This method would overestimate the volume density in clusters where the stars
are uniformly distributed over the line of sight, but would provide a more
accurate measurement of the volume density toward observed surface density
peaks resulting from a sharply peaked or clumpy volume density field.  The
method is particularly applicable when the number of stars in a volume density
peak is larger than the number of surrounding clusters stars in the line of
sight.  This method is also more applicable in highly elongated or filamentary
clusters.  To minimize the effect of field star contamination, we apply this
method only in regions were the surface density of cluster members is much
larger than the density of field stars.  The reported densities are only
densities of detected stars; stars too reddened or too dim to detect in our
data are not taken into account.

\section{Results\label{results}}

\subsection{GGD~12-15}

\begin{figure}
\epsscale{1}
\plotone{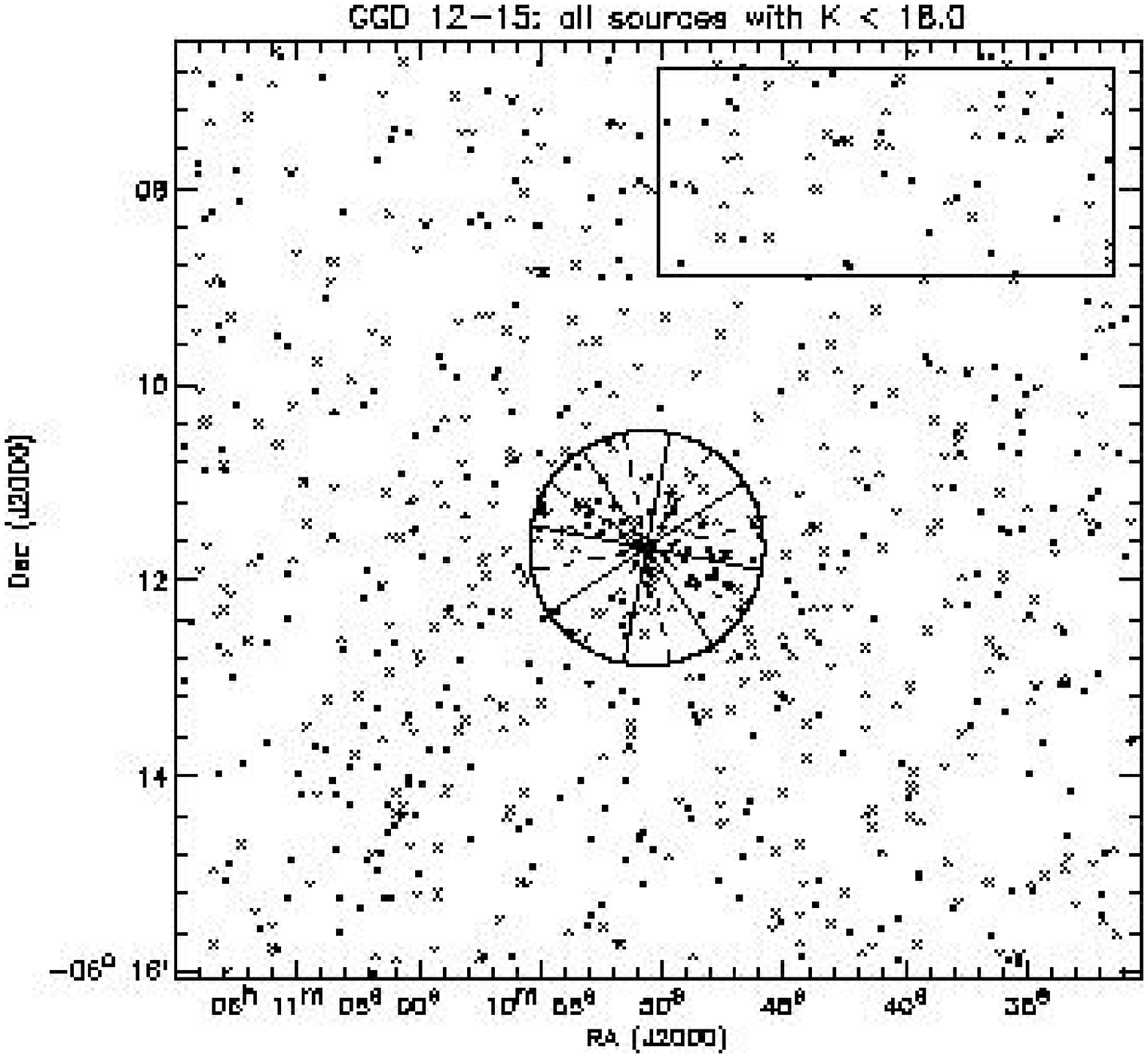}
\caption{Plot of all $K$--band detections brighter than $K = 18$ in GGD~12-15.
The box marks the off-field used for characterization of local field
star contamination.  The circle marks the cluster region boundary as determined
in Fig.~\ref{ggd1215rad}  (see Section~\ref{radial}).  The overlapping
$45^{\circ}$ position angle bins used to make the Nyquist sampled azimuthal
distribution histograms in Fig.~\ref{ggd1215az} are marked by the alternating
black and gray wedge outlines.\label{ggd1215regions}}
\end{figure}

\begin{figure}
\epsscale{1}
\plotone{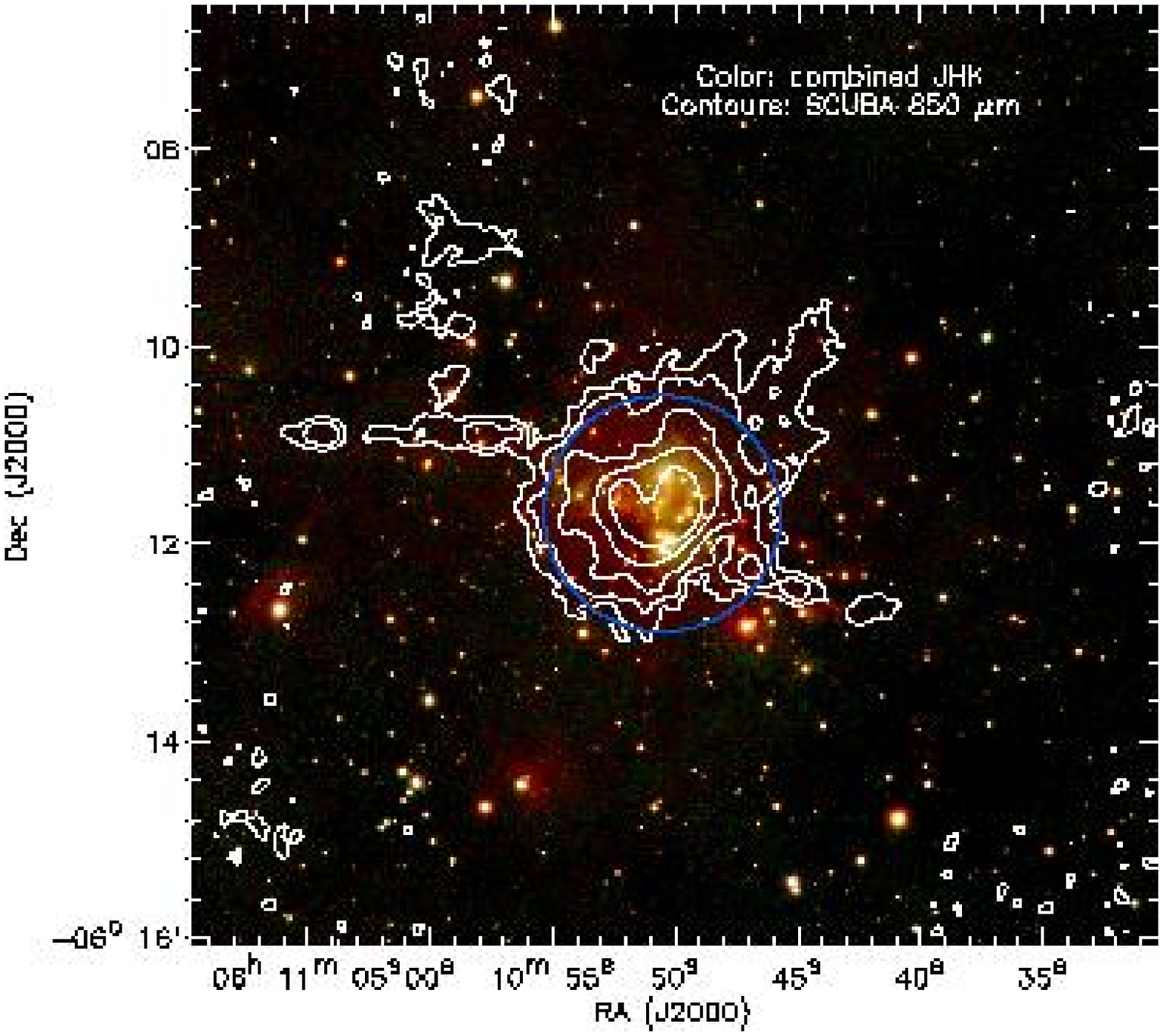}
\caption{$JHK$ log--scaled color image of the GGD~12-15 region from FLAMINGOS
on MMT overlaid with 850~$\mu$m dust emission contours from SCUBA on JCMT.  The
contour levels start at 1.75~mJy~pixel$^{-1}$ 
($2^{\prime\prime} \times 2^{\prime\prime}$~pixels), with successive contour
levels at double the previous level.  The blue circle shows the cluster region
boundary determined in Fig.~\ref{ggd1215rad}.\label{ggd1215jhks}}
\end{figure}

\begin{figure}
\epsscale{1}
\plotone{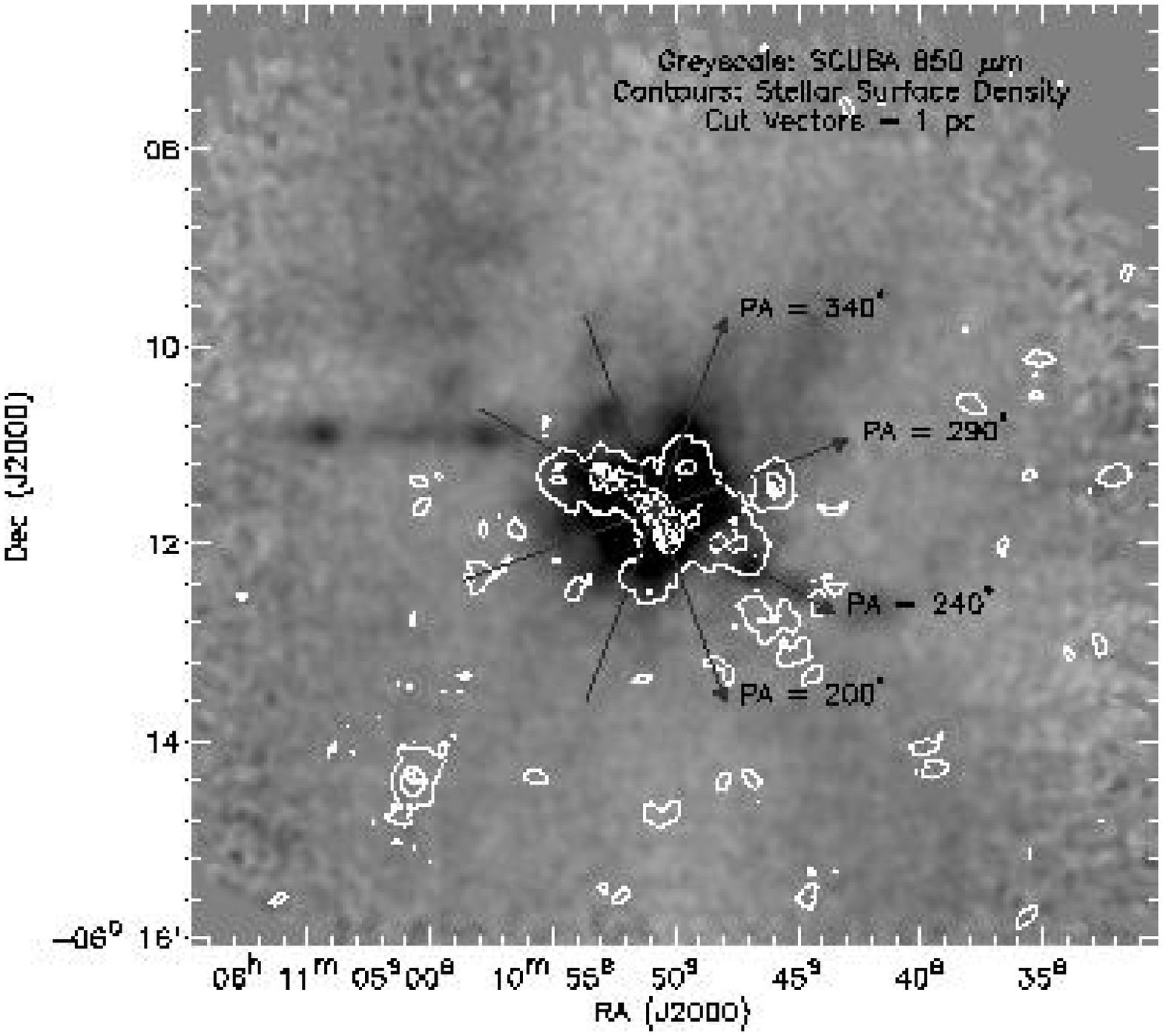}
\caption{850~$\mu$m SCUBA map of GGD~12-15 in negative grayscale, scaled 
to show faint filamentary structure.  Stellar surface density contours measured from all $K$--band point sources brighter than $K = 18$ are overlaid. 
Contours begin at 435~pc$^{-2}$ ($5\sigma$ above median 
field star density), and increase at an interval of 750~pc$^{-2}$.  The 
vectors denote the location and extent of the stellar surface density map cuts
plotted in Fig.~\ref{ggd1215slices}.  Each vector is 1~pc in length at the 
assumed distance of 830~pc, approximately the distance a star moving
$1 km s^{-1}$ will travel in 1~Myr.\label{ggd1215sg}}
\end{figure}

\begin{figure}
\epsscale{1}
\plotone{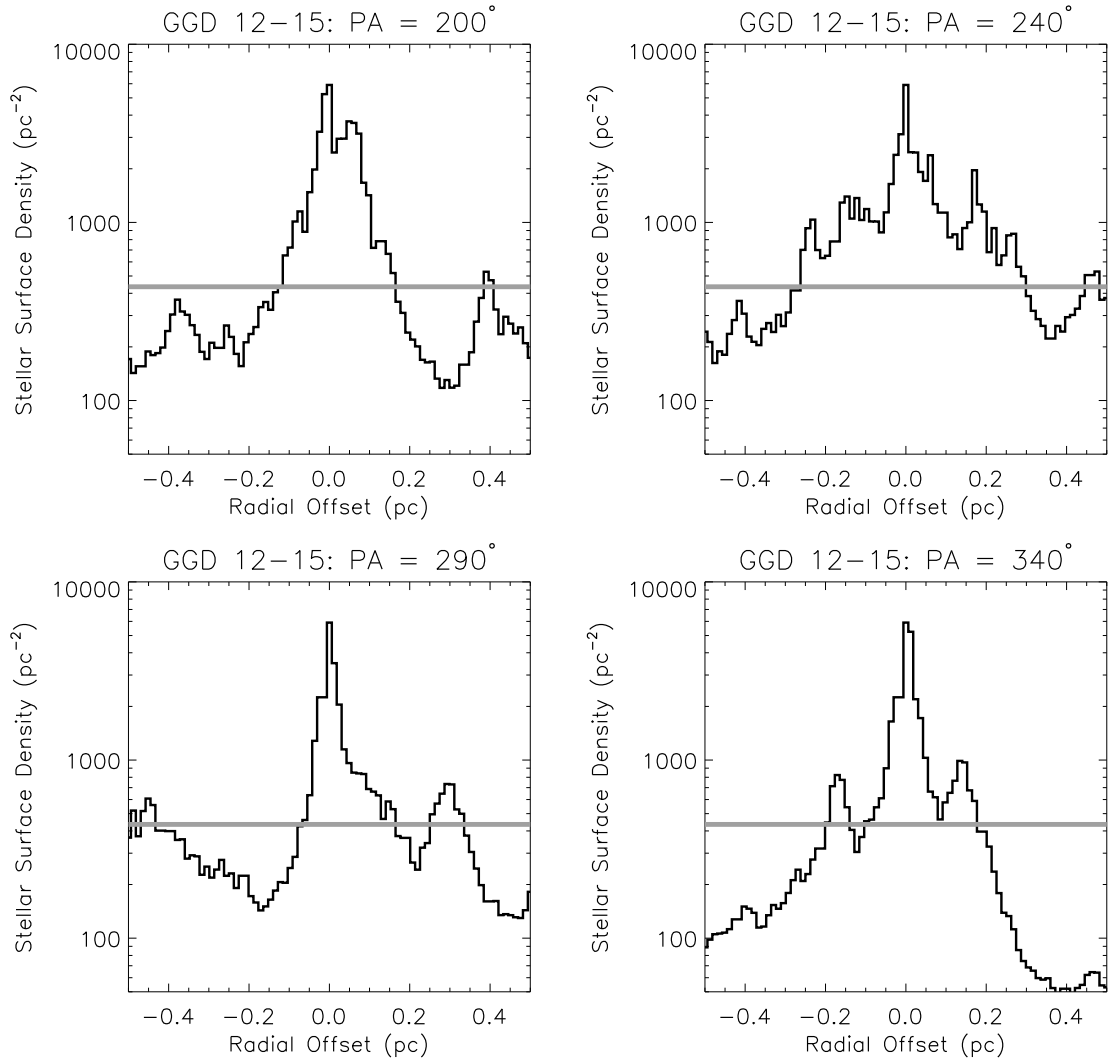}
\caption{The black lines in the above plots are stellar surface density map 
cuts for GGD~12-15.  The vectors plotted in Fig.~\ref{ggd1215sg} mark the 
position, orientation, and direction of increasing offset from cluster center.
Note the strong central peak, multiple sub-peaks, and lack of uniform circular
symmetry.  The gray line marks the $5\sigma$ above median field star density 
level for comparison, which corresponds to the first contour level in 
Fig.~\ref{ggd1215sg}.\label{ggd1215slices}}
\end{figure}

By inspection of the contamination--subtracted KMH of the cluster region of
GGD~12-15 (Fig.~\ref{ggd1215klf}) it is clear that the cluster population 
dominates field star counts at all magnitudes throughout the sensitivity
range of our data.  Hence we choose to use a limiting magnitude equal to the 
end of the first complete magnitude bin brighter than the 90\% completeness
limit, $K = 18$.  The contamination--subtracted radial density profile fit 
(Fig.~\ref{ggd1215rad}) yields a cluster core radius of 0.24~pc, but the large
variation in the azimuthal distribution histogram (Fig.~\ref{ggd1215az}) 
suggests significant asymmetric structure in the cluster core.  Indeed, there
are two peaks in the azimuthal distribution which are separated by
$180^{\circ}$, a signature of linear structure.  This is 
confirmed by inspection of the stellar distribution (see
Figs.~\ref{ggd1215regions}~and~\ref{ggd1215jhks}), the stellar density map
(Fig.~\ref{ggd1215sg}), and the stellar density map slice plots 
(Fig.~\ref{ggd1215slices}).
Finally, the contamination--subtracted $AAP$ of GGD~12-15 is 1.588,
$3.77\sigma$ above the simulated mean $AAP$, suggesting only a 0.008\% 
probability that the radially averaged azimuthal distribution observed is
consistent with a circularly symmetric configuration.  Clearly, an azimuthally
averaged profile is a poor characterization for the highly structured nature of
this cluster.

The refined cluster core center point approximately coincides with the location
of peak density of the stellar density distribution map of GGD~12-15, showing 
that there is a well-defined and cohesive stellar surface density peak at the 
center of the cluster core, even though there is significant structure 
evident in the stellar density map.  
The position of peak stellar density occurs between two peaks in the core dust
emission.  This suggests either that the central stars
are in the process of opening a cavity in the cloud core or that local 
extinction has in fact drastically reduced our sensitivity to stars 
in the areas of high dust density.  
Furthermore, we report the discovery of 850~$\mu$m dust emission filaments (see
Fig.~\ref{ggd1215jhks}) extending from the main core which has been previously
investigated by \citet{lhd90}.  The orientation of these filaments is nearly
identical to the orientation of the elongated stellar distribution (see
Fig.~\ref{ggd1215sg}).  
We argue that this is evidence that star formation in GGD~12-15 is primarily
occuring along the molecular cloud filament and that the stars have not had
adequate time to migrate significantly from their birth sites.

\subsection{IRAS~20050+2720\label{iras20050}}

\begin{figure}
\epsscale{0.8}
\plotone{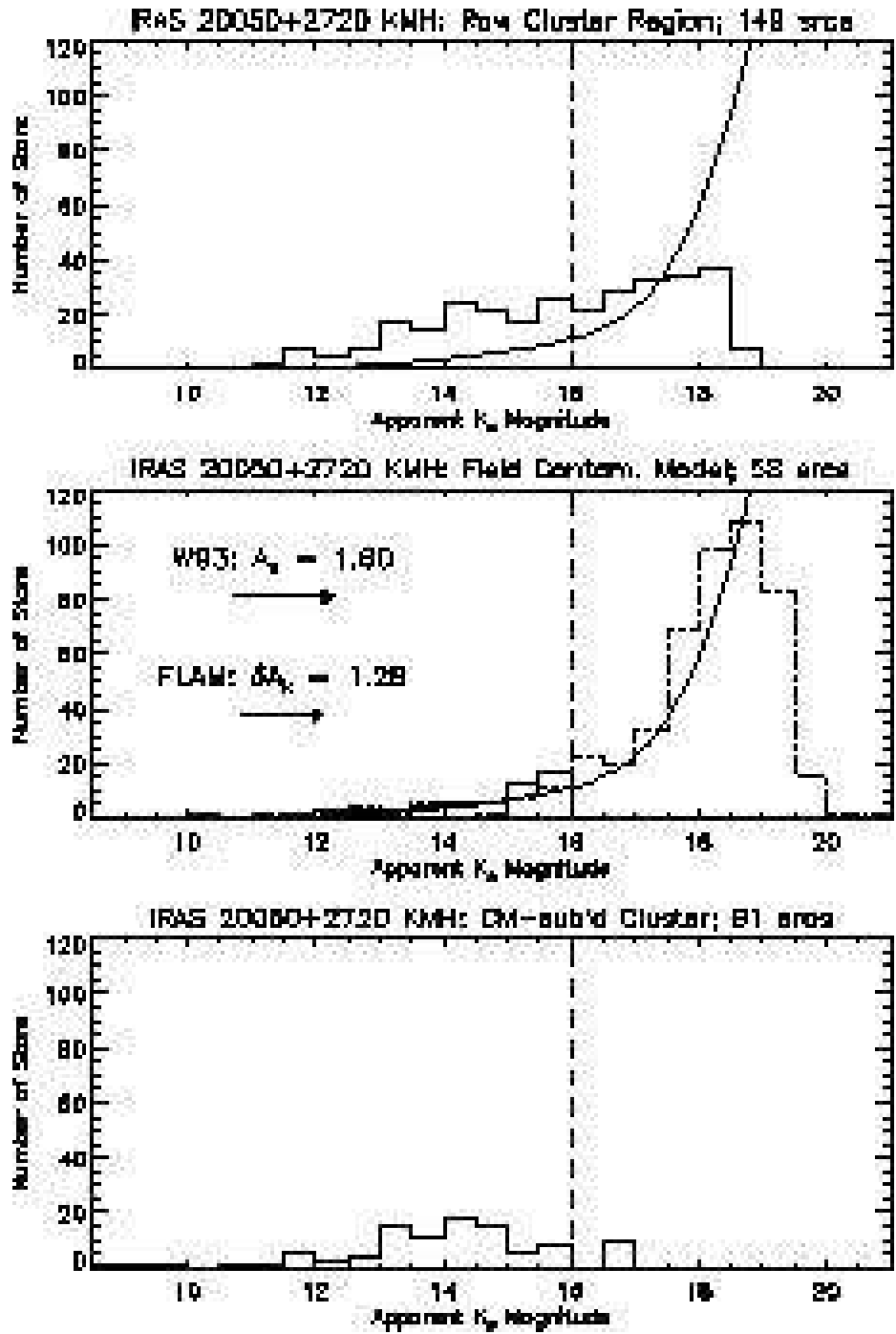}
\caption{$K_{s}$--band apparent magnitude histograms for IRAS~20050+2720.  For
a detailed description, see Fig.~\ref{ggd1215klf}.\label{iras20050klf}}
\end{figure}

\begin{figure}
\epsscale{1}
\plotone{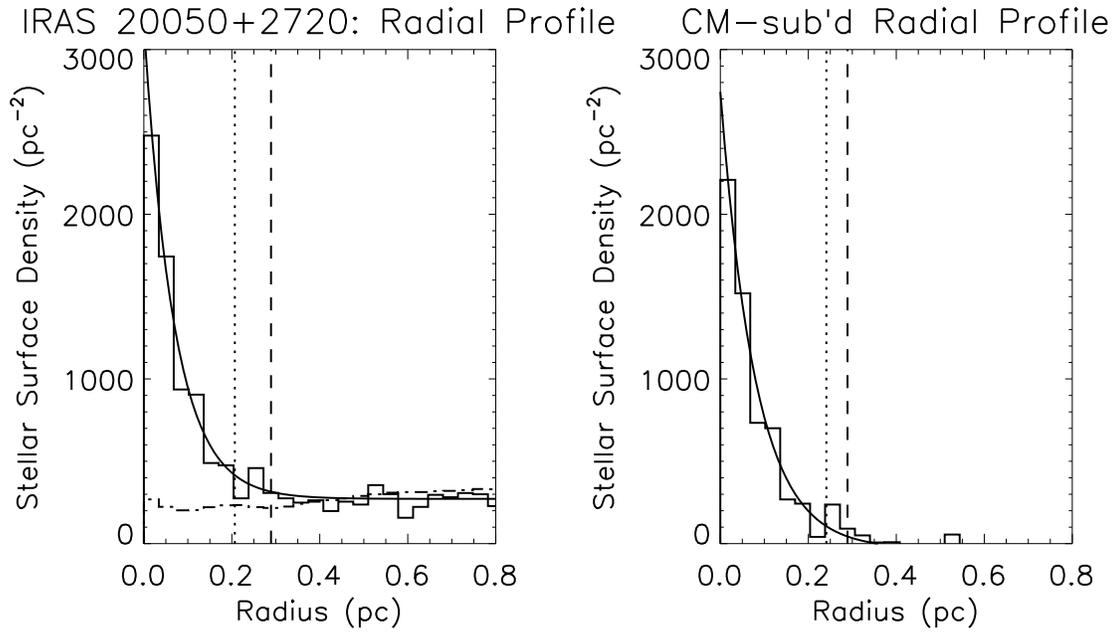}
\caption{Azimuthally averaged stellar radial density profiles for the
IRAS~20050+2720 region.  For a detailed description, see
Fig.~\ref{ggd1215rad}.\label{iras20050rad}}
\end{figure}

\begin{figure}
\epsscale{1}
\plotone{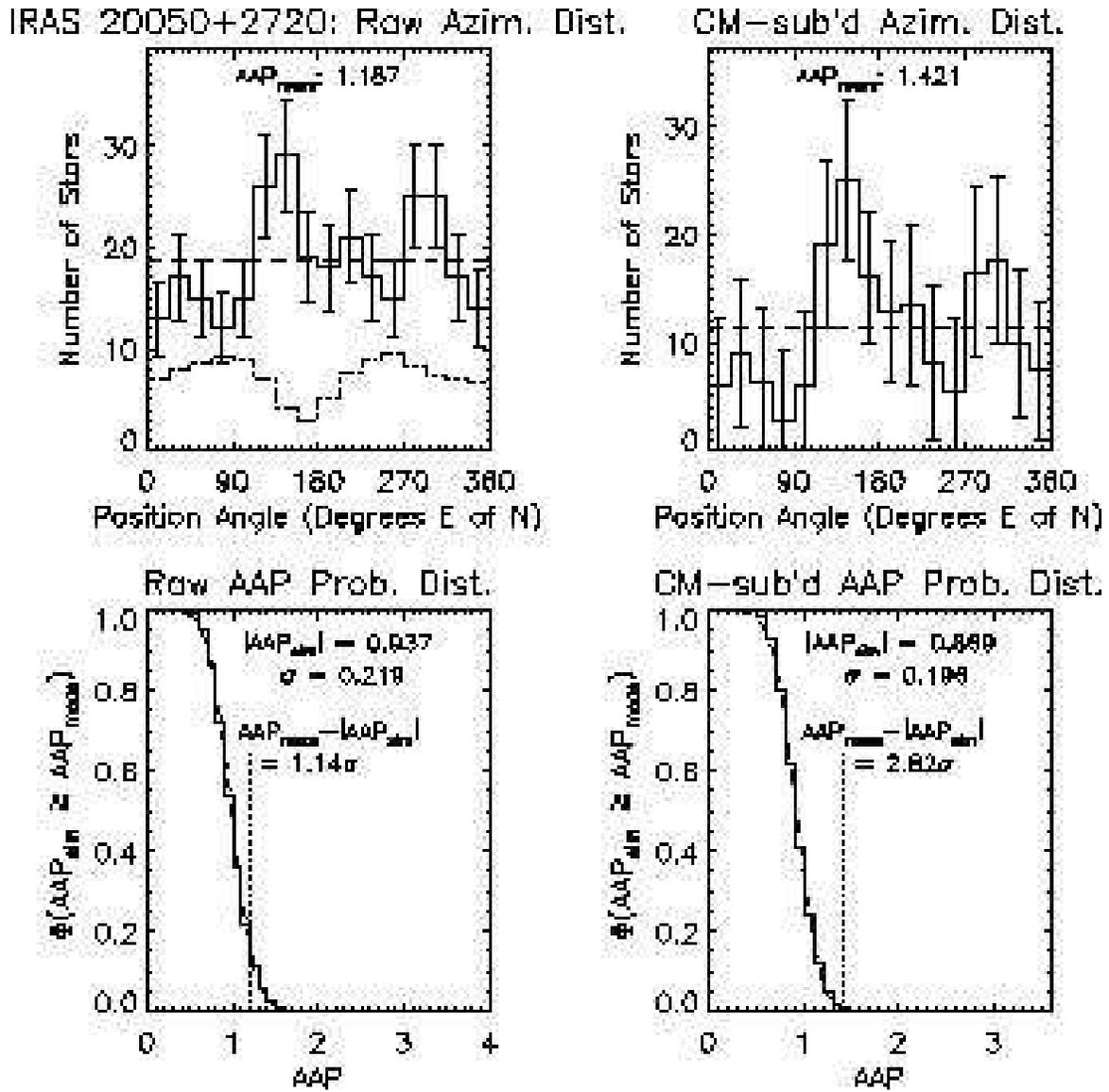}
\caption{Azimuthal distribution histograms and $AAP$ probability distribution
histograms for IRAS~20050+2720.  For a detailed description, see
Fig.~\ref{ggd1215az}.\label{iras20050az}}
\end{figure}

\begin{figure}
\epsscale{1}
\plotone{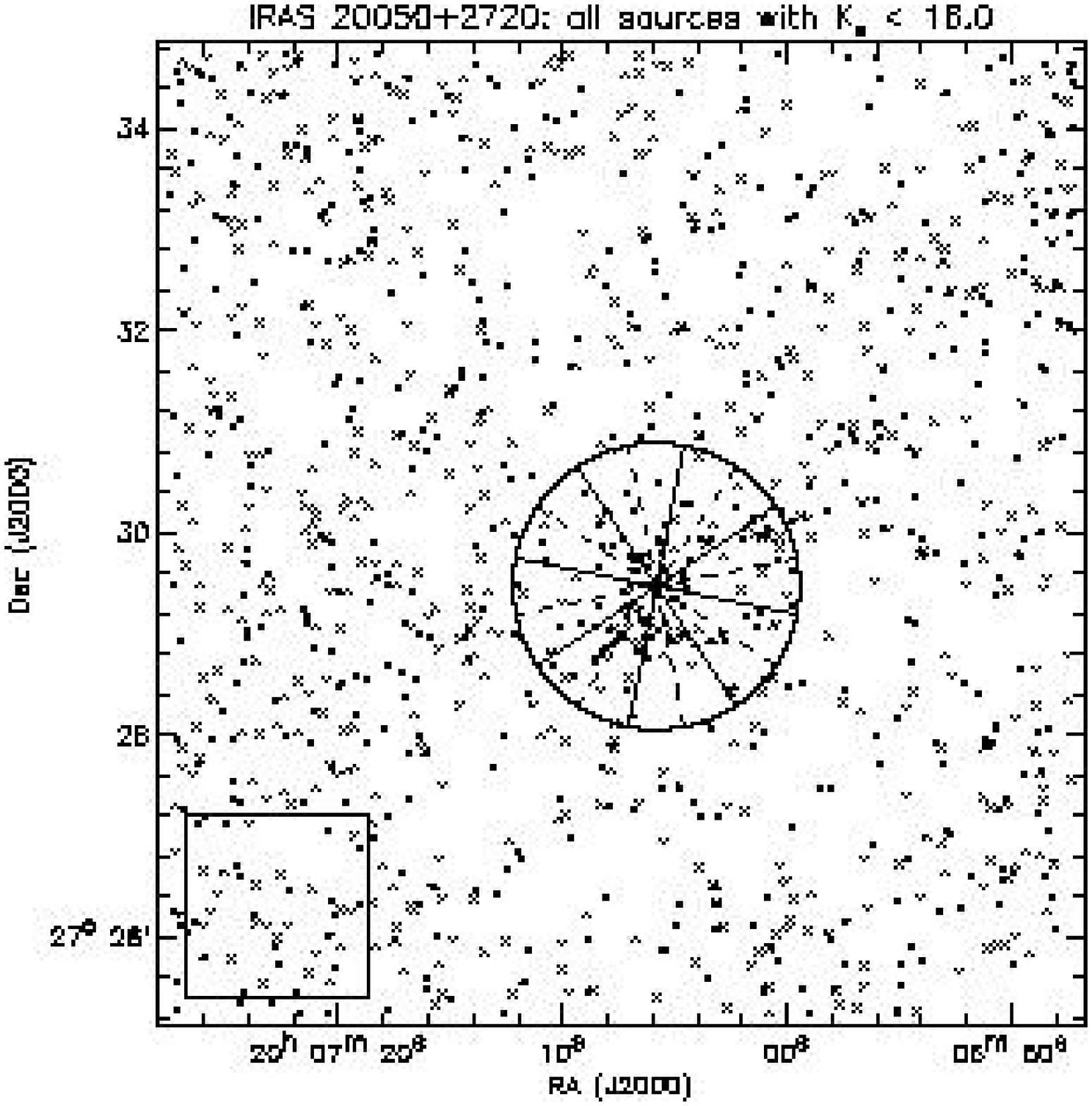}
\caption{Plot of all $K_{s}$--band detections brighter than $K_{s} = 16$ in 
IRAS~20050+2720.  
The box marks the off-field used for characterization of local field
star contamination.  The circle marks the cluster region boundary as determined
in Fig.~\ref{iras20050rad}  (see Section~\ref{radial}).  The overlapping
$45^{\circ}$ positional angle bins used to make the Nyquist sampled azimuthal
distribution histograms in Fig.~\ref{iras20050az} are marked by the alternating
black and gray wedge outlines.\label{iras20050regions}}
\end{figure}

\begin{figure}
\epsscale{1}
\plotone{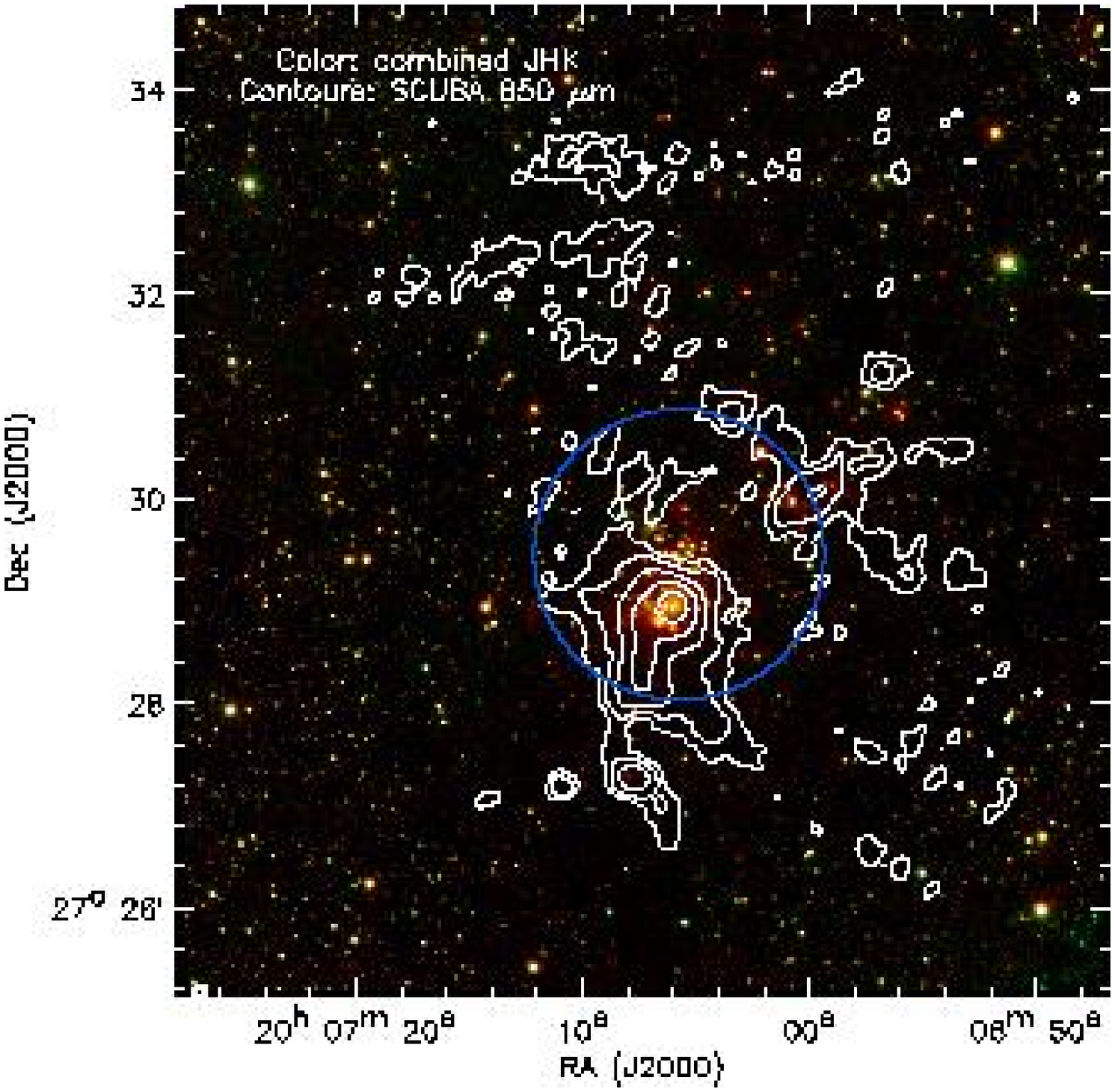}
\caption{$JHK_{s}$ log--scaled color image of the IRAS~20050+2720 region from
FLAMINGOS on MMT overlaid with 850~$\mu$m dust emission contours from SCUBA on
JCMT. The contour levels start at 1.75~mJy~pixel$^{-1}$
($2^{\prime\prime} \times 2^{\prime\prime}$~pixels), with successive contour
levels at two times the previous level.  The blue circle shows the cluster
region boundary determined in Fig.~\ref{iras20050rad}.\label{iras20050jhks}}
\end{figure}

\begin{figure}
\epsscale{1}
\plotone{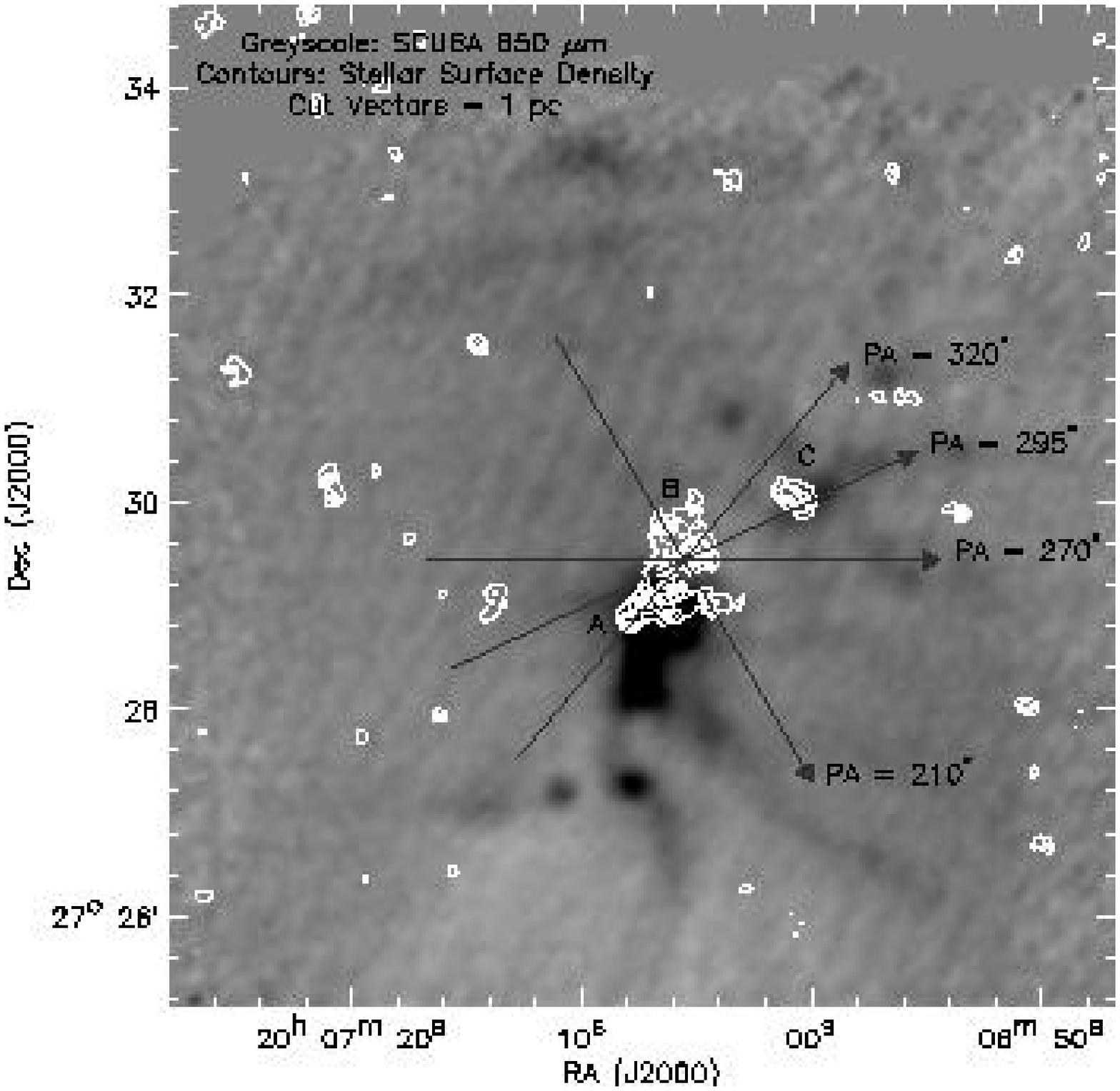}
\caption{850~$\mu$m SCUBA map of IRAS~20050+2720 in negative grayscale, scaled
to show faint filamentary structure.  Stellar surface density contours measured
from all $K_{s}$--band point sources brighter than $K_{s} = 16$ are overlaid.  
Contours begin at 1450~pc$^{-2}$ ($5\sigma$ above median 
field star density), and increase at an interval of 750~pc$^{-2}$.  The 
vectors denote the location and extent of the stellar density map cuts plotted
in Fig.~\ref{iras20050slices}.  Each vector is 1~pc in length at the assumed
distance of 700~pc, approximately the distance a star moving $1 km s^{-1}$ will
travel in 1~Myr.  The letters A, B, and C denote the contour groups that 
correspond to the three subclusters given those names by
\citet{chen97}.\label{iras20050sg}}
\end{figure}

\begin{figure}
\epsscale{1}
\plotone{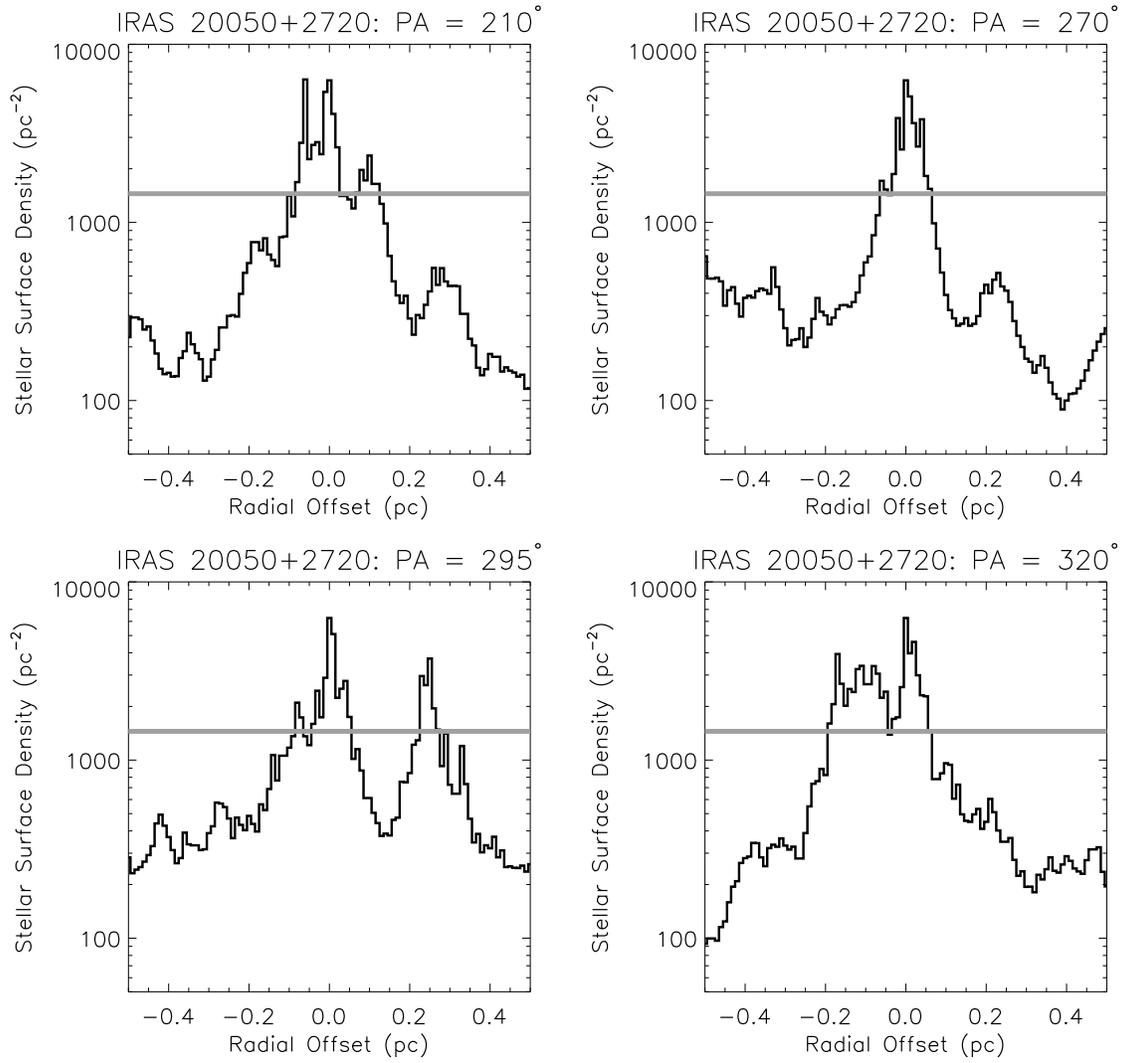}
\caption{The black lines in the above plots are stellar surface density map 
cuts for IRAS~20050+2720.  The vectors plotted in Fig.~\ref{iras20050sg} mark 
the position, orientation, and direction of increasing offset from cluster 
center.  Note the multiple strong peaks and deviation from circular symmetry. 
The gray line marks the $5\sigma$ above median field star density level for 
comparison, which corresponds to the first contour level in 
Fig.~\ref{iras20050sg}.\label{iras20050slices}}
\end{figure}

The contamination--subtracted KMH of the cluster region of the IRAS~20050+2720 
field (Fig.~\ref{iras20050klf}) shows that the cluster population becomes
dominated by field star contamination for sources fainter than $K_{s} = 16$,
hence we adopt that as our limiting magnitude for this work.  The radial
density profile fit (Fig.~\ref{iras20050rad}) yields a cluster core radius of
0.24~pc, but as with GGD~12-15, there is significant evidence to suggest that
this is a very poor characterization due to significant deviation from circular 
symmetry \citep[see Figs.~\ref{iras20050regions}~and~\ref{iras20050jhks}, and
note that][reported clear evidence of three distinct subclusters in
IRAS~20050+2720, which they designated subclusters~A,~B,~and~C.]{chen97}. 
The refined cluster core center point of IRAS~20050+2720 is not located at the
position of peak density in the stellar density map, but is instead located
approximately $20^{\prime\prime}$ to the southwest.  The azimuthal distribution
histogram (Fig.~\ref{iras20050az}) has an obvious peak corresponding to the
contribution from subcluster~A and a less prominent one corresponding to
subcluster~C (see Figs.~\ref{iras20050sg}~and~\ref{iras20050slices}).  
The contamination--subtracted $AAP$ of IRAS~20050+2720 is 1.421, $2.82\sigma$
above the simulated mean $AAP$, suggesting a $0.2\%$ probability that the
azimuthal distribution observed is consistent with a circularly symmetric
confiuration.  
It is clear that extended stellar distributions with subclustering to this
degree are poorly described using methods that utilize azimuthal averaging.  

While the locations of peak stellar surface density and the refined cluster
core center point are not coincident, both positions are within subcluster~B as
defined by \citet{chen97}.  Subcluster~B is in a region lacking 850~$\mu$m
emission within an otherwise apparently colinear filamentary structure in the
dust emission map (see Fig.~\ref{iras20050jhks}).  This subcluster has 
significantly less reddening, dispersion in $H-K$, and associated 
$K$--band nebulosity as compared to subclusters~A~and~C \citep{chen97}, which
are both located in areas of significant 850~$\mu$m emission.  However, the
stellar surface density map (see Fig.~\ref{iras20050sg}) does not show a
clear boundary between subclusters~A~and~B, leaving open the possibility that
these two subclusters are not distinct structures, but instead may be
subregions of a single subcluster that have very different amounts of
associated dust in the line of sight.  

Our wide-field 850~$\mu$m map is nearly identical in morphology to the 1.3~mm
map presented in \citet{chin01}, including the small cores named MM2 and MM3
located south of the cluster region and the small unnamed core to the northwest
of the cluster.  Given the lack of dust emission in the vicinity of
subcluster~B and the multiple high velocity molecular outflows detected in this
region, our analysis suggests that it has very recently dispersed its natal
molecular gas while star formation proceeds along the rest of the filament, in
agreement with the conclusions reached by \citet{chen97}. 

\subsection{NGC~7129\label{ngc7129}}

\begin{figure}
\epsscale{0.8}
\plotone{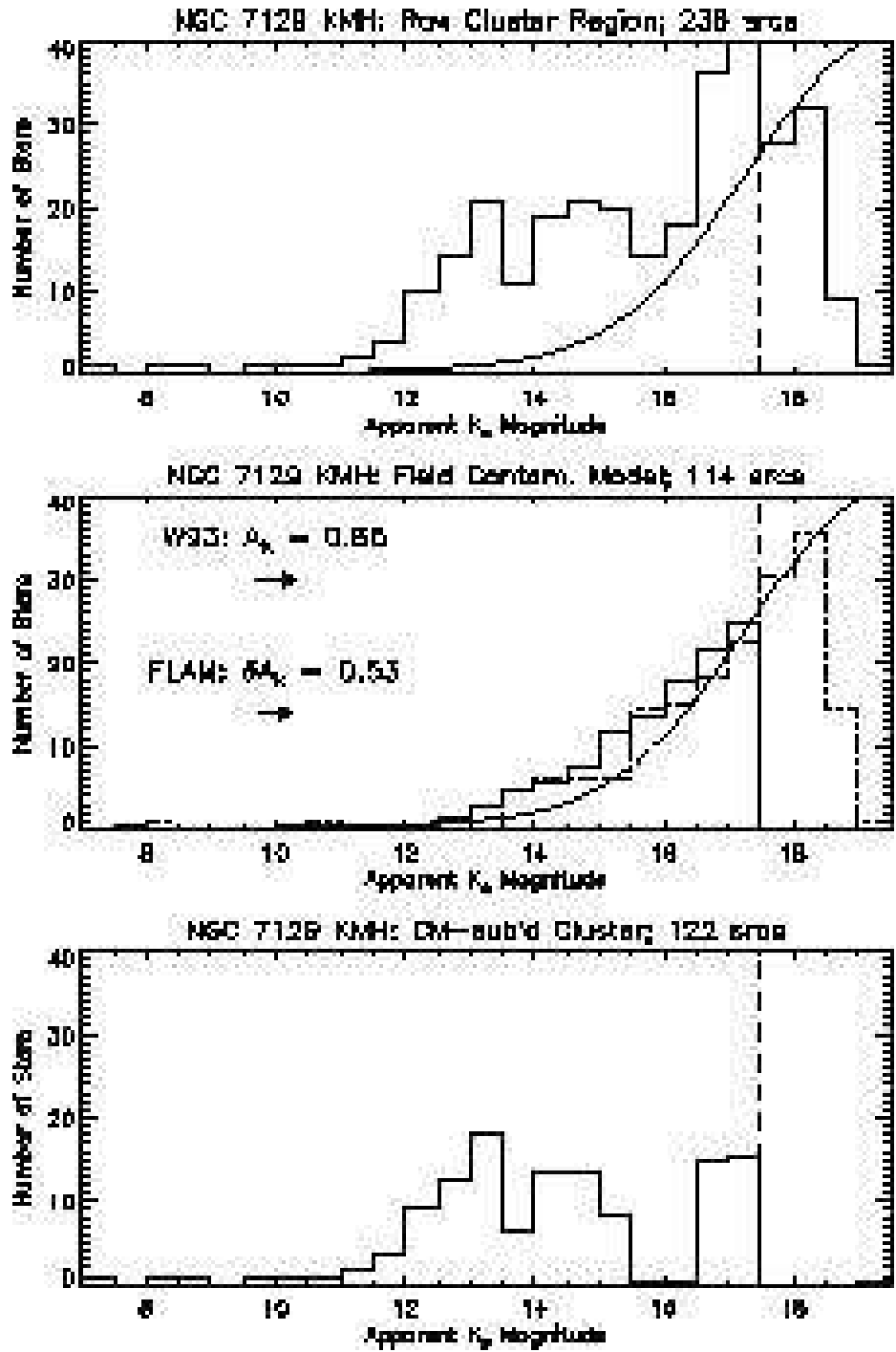}
\caption{$K_{s}$--band apparent magnitude histograms for NGC~7129.  For
a detailed description, see Fig.~\ref{ggd1215klf}.\label{ngc7129klf}}
\end{figure}

\begin{figure}
\epsscale{1}
\plotone{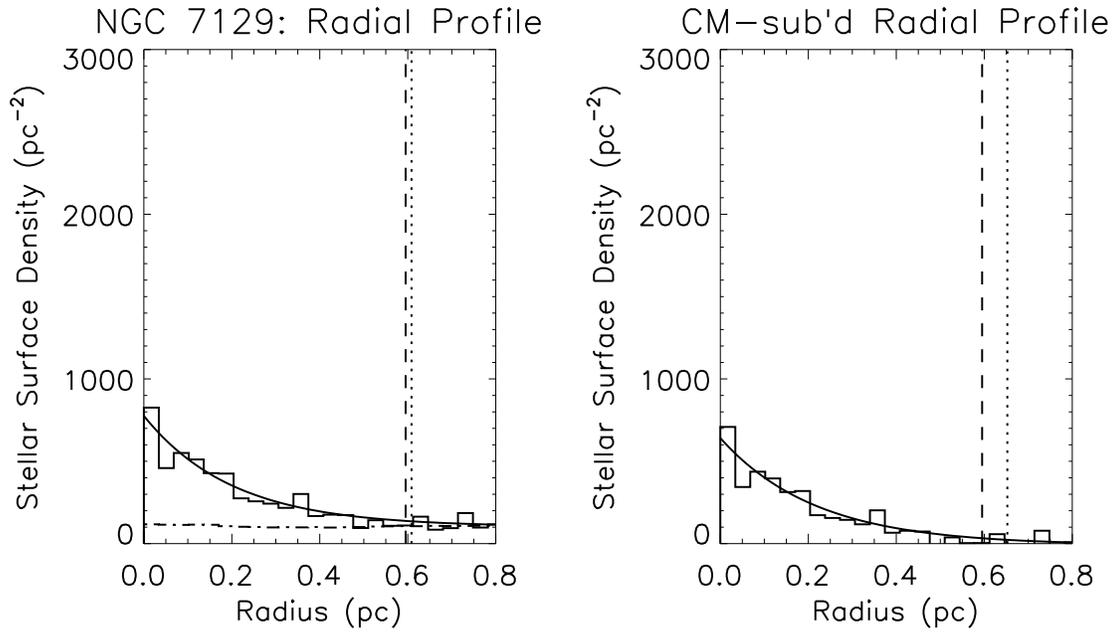}
\caption{Azimuthally averaged stellar radial density profiles for the NGC~7129
region.  For a detailed description, see
Fig.~\ref{ggd1215rad}.\label{ngc7129rad}}
\end{figure}

\begin{figure}
\epsscale{1}
\plotone{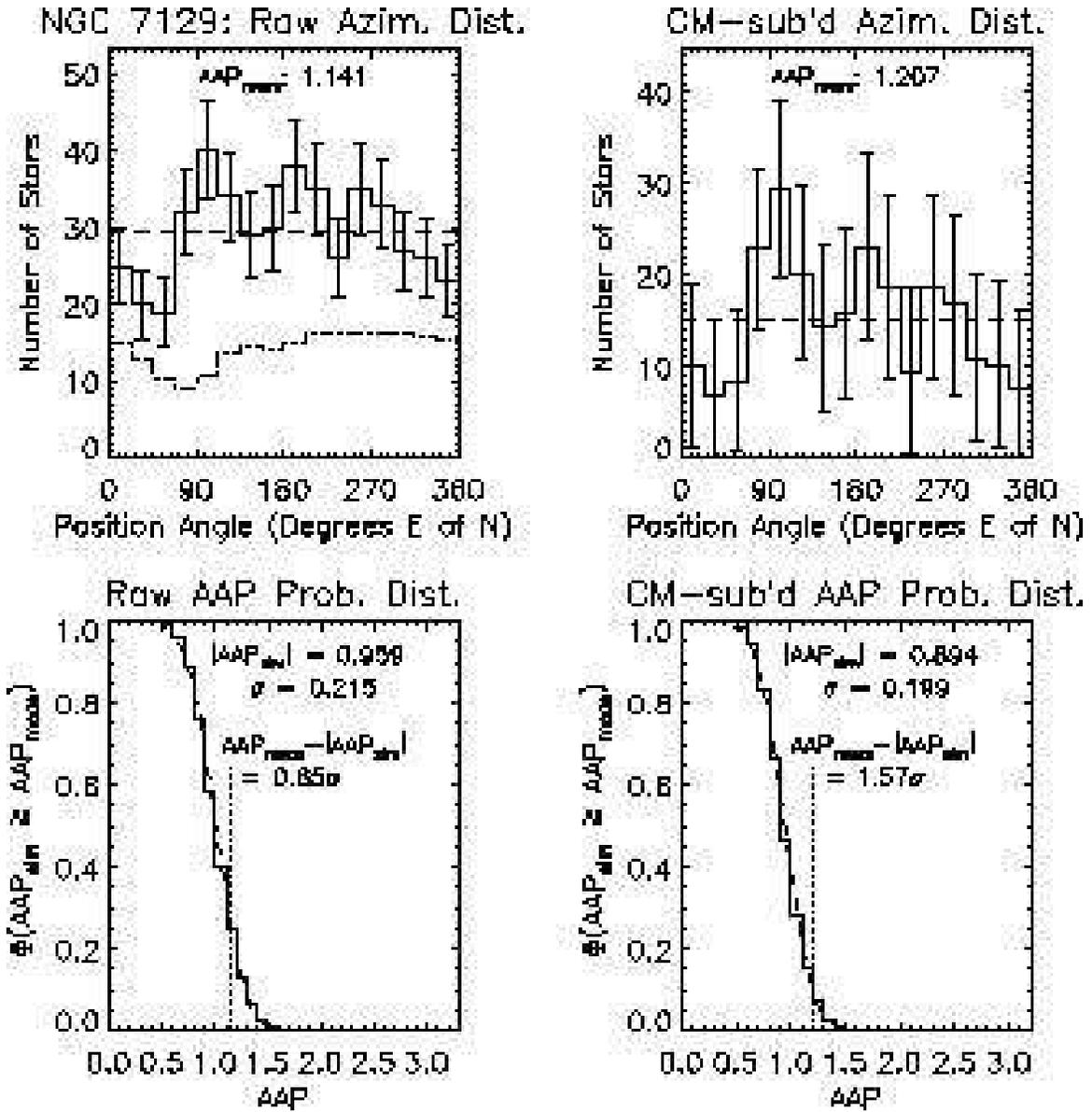}
\caption{Azimuthal distribution histograms and $AAP$ probability distribution
histograms for NGC~7129.  For a detailed description, see
Fig.~\ref{ggd1215az}.\label{ngc7129az}}
\end{figure}

\begin{figure}
\epsscale{1}
\plotone{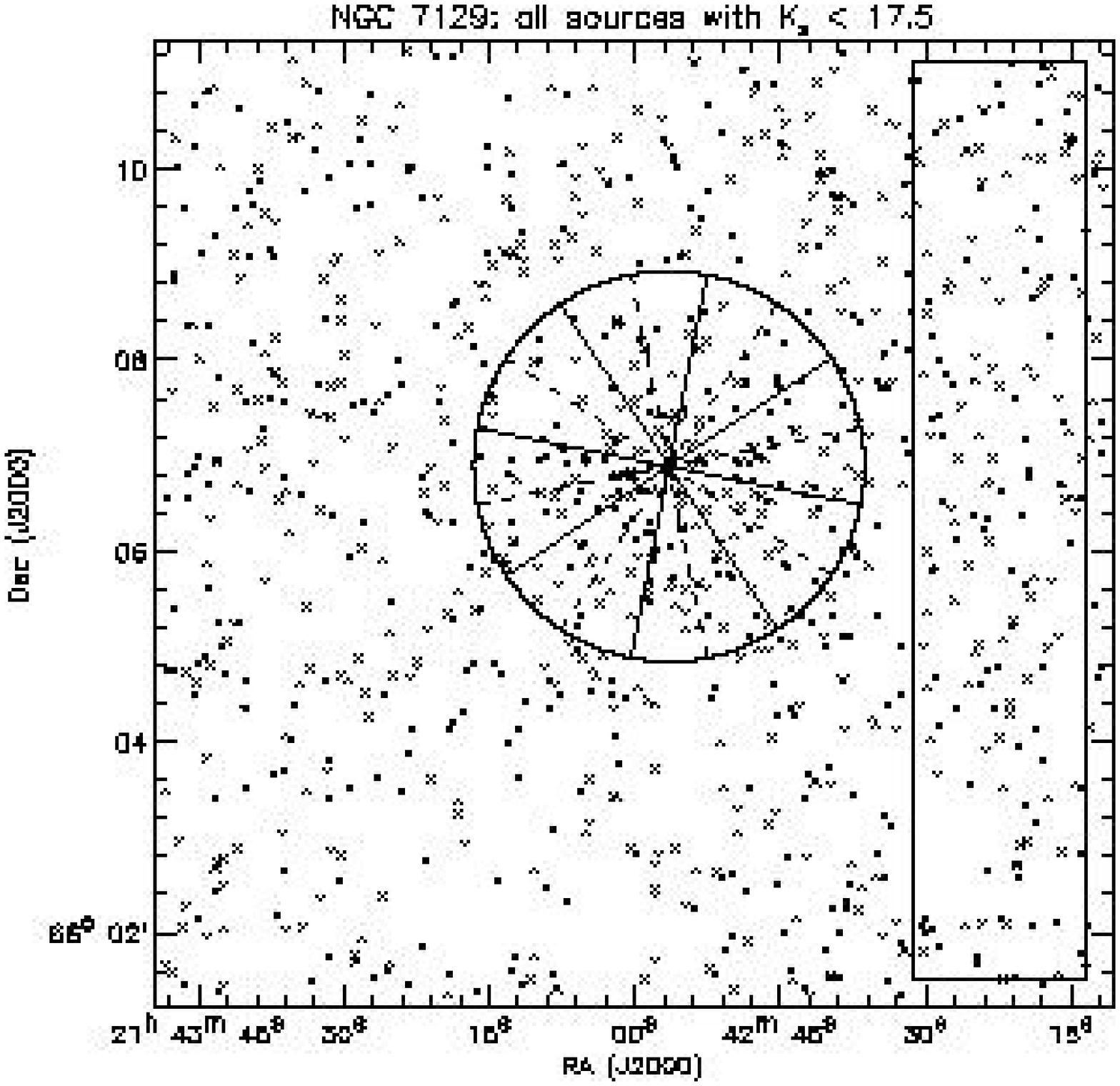}
\caption{Plot of all $K_{s}$--band detections brighter than $K_{s} = 17.5$ in 
NGC~7129.  
The box marks the off-field used for characterization of local field
star contamination.  The circle marks the cluster region boundary as determined
in Fig.~\ref{ngc7129rad}  (see Section~\ref{radial}).  The overlapping
$45^{\circ}$ positional angle bins used to make the Nyquist sampled azimuthal
distribution histograms in Fig.~\ref{ngc7129az} are marked by the alternating
black and gray wedge outlines.\label{ngc7129regions}}
\end{figure}

\begin{figure}
\epsscale{1}
\plotone{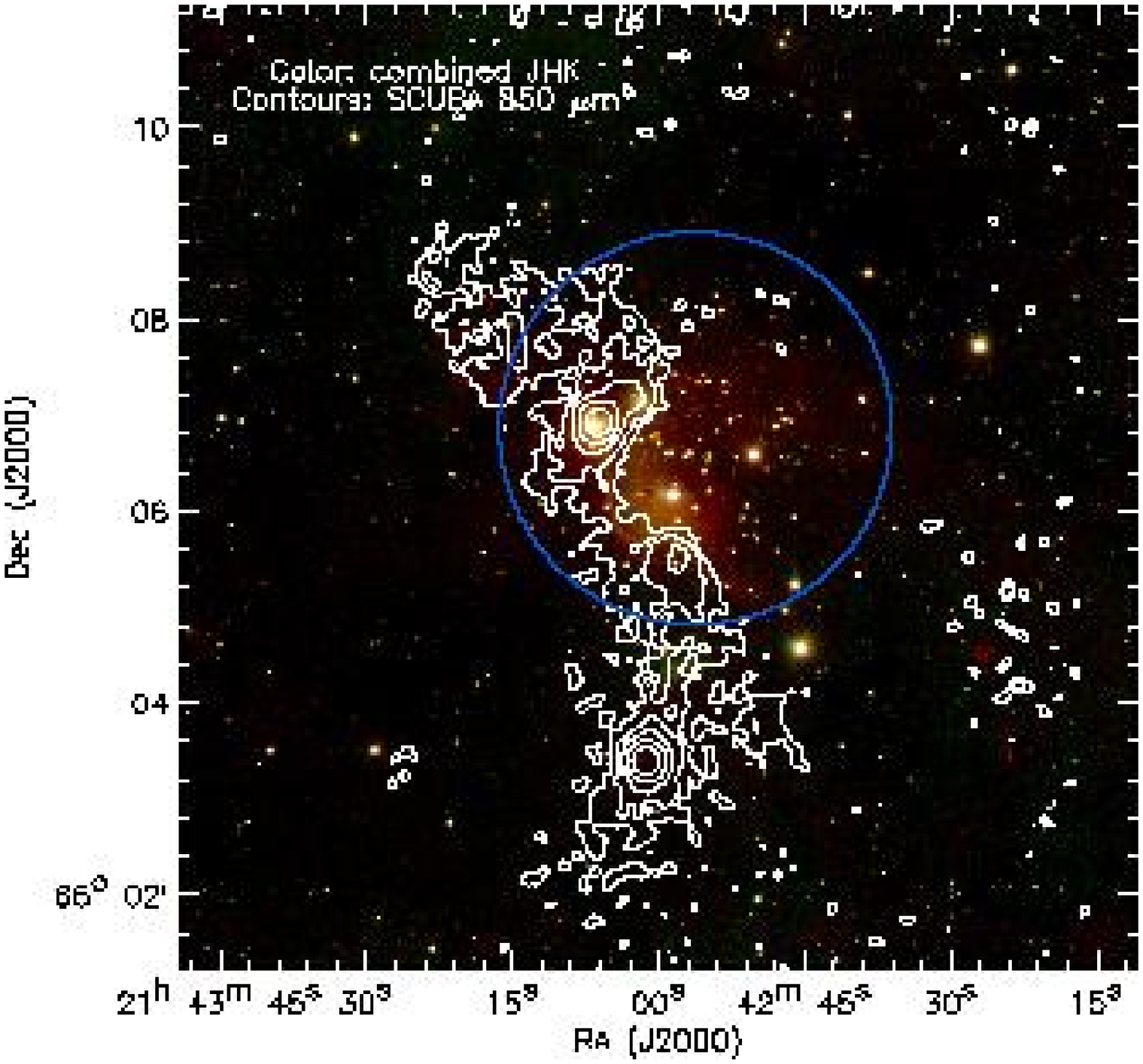}
\caption{$JHK_{s}$ log--scaled color image of the NGC~7129 region from
FLAMINGOS on MMT overlaid with 850~$\mu$m dust emission contours from SCUBA on
JCMT.  The contour levels start at 1.75~mJy~pixel$^{-1}$
($2^{\prime\prime} \times 2^{\prime\prime}$~pixels), with successive contour
levels at double the previous level.  The blue circle shows the cluster region
boundary determined in Fig.~\ref{ngc7129rad}.\label{ngc7129jhks}}
\end{figure}

By inspection of the contamination--subtracted KMH of the cluster region of 
NGC~7129 (Fig.~\ref{ngc7129klf}) we argue that the cluster population 
dominates field star counts at most magnitudes throughout the
sensitivity range of our data.  Hence we choose to use a limiting magnitude
equal to the end of the first complete magnitude bin brighter than the
90\% completeness limit, $K = 17.5$.  
The contamination--subtracted
radial density profile fit (Fig.~\ref{ngc7129rad}) yields a cluster core radius
of 0.65~pc, much larger than the that of the other two clusters presented in
this work.  Also, the azimuthal distribution histogram (Fig.~\ref{ngc7129az})
is quite featureless by comparison to the other two clusters presented,
suggesting that the cluster is approximately circularly symmetric.  This can
also be seen by inspection of the stellar distribution
(Fig.~\ref{ngc7129regions}) which is clearly located primarily in the cavity
defined by the 850~$\mu$m emission (see Fig.~\ref{ngc7129jhks}).  
Finally, the contamination--subtracted $AAP$ of NGC~7129 is 1.207, $1.57\sigma$
above the simulated mean $AAP$, suggesting a 5.8\% probability that the
azimuthal distribution observed is consistent with a circularly symmetric
configuration.  While this result suggests some asymmetry may be present,
inconsistency with a circularly symmetric distribution is not statistically
significant.

\begin{figure}
\epsscale{1}
\plotone{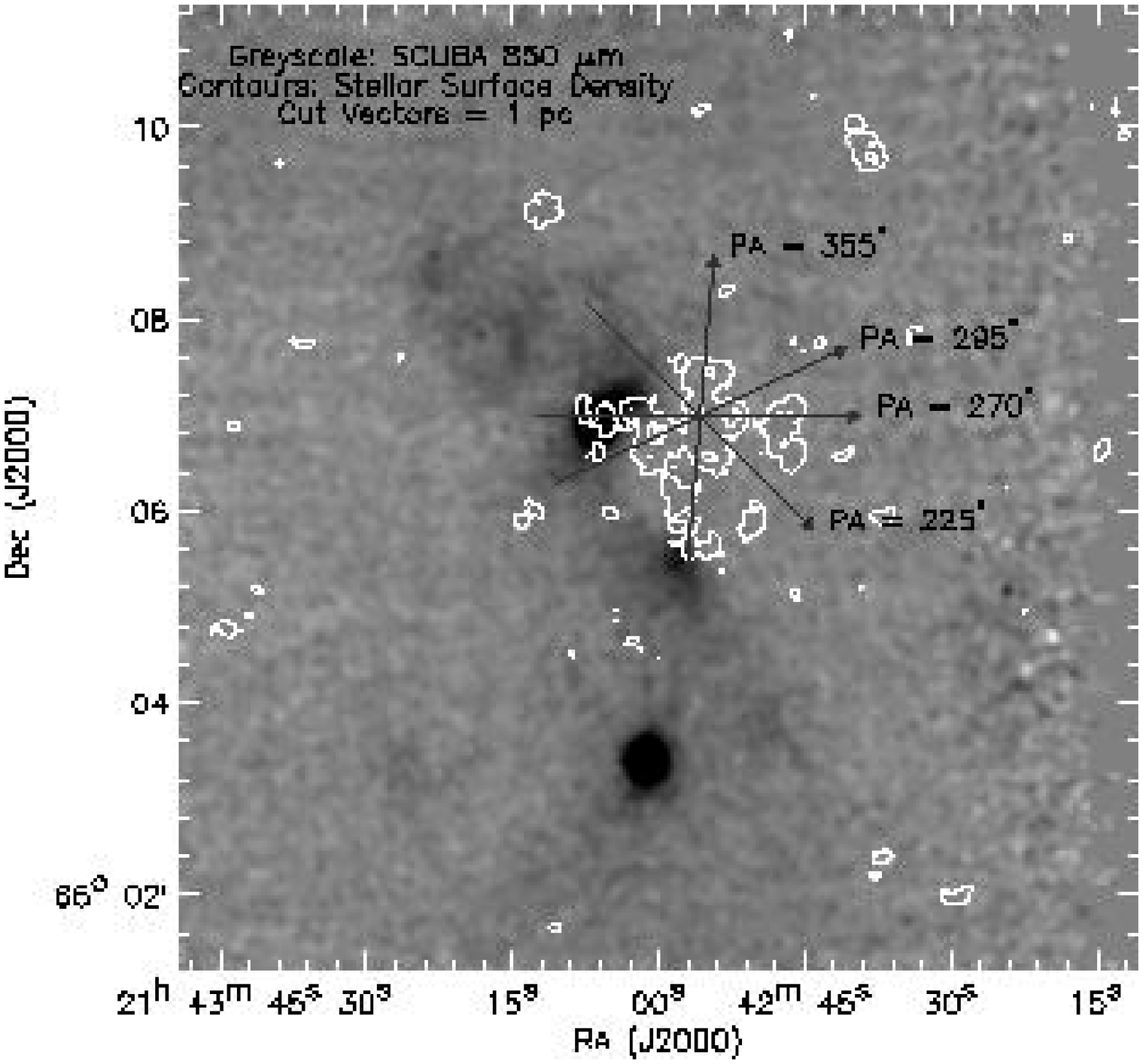}
\caption{850~$\mu$m SCUBA map of NGC~7129 in negative grayscale, scaled to show
faint filamentary structure.  Stellar surface density contours measured from
all $K_{s}$-band point sources brighter than $K_{s} = 17.5$ are overlaid. 
Contours begin at 475~pc$^{-2}$ ($5\sigma$ above median 
field star density), and increase at an interval of 750~pc$^{-2}$.  The 
vectors denote the location and extent of the  stellar density map cuts 
plotted in Fig.~\ref{ngc7129slices}.  Each vector is 1~pc in length at the 
assumed distance of 1~kpc, approximately the distance a star moving
$1 km s^{-1}$ will travel in 1~Myr.\label{ngc7129sg}}
\end{figure}

\begin{figure}
\epsscale{1}
\plotone{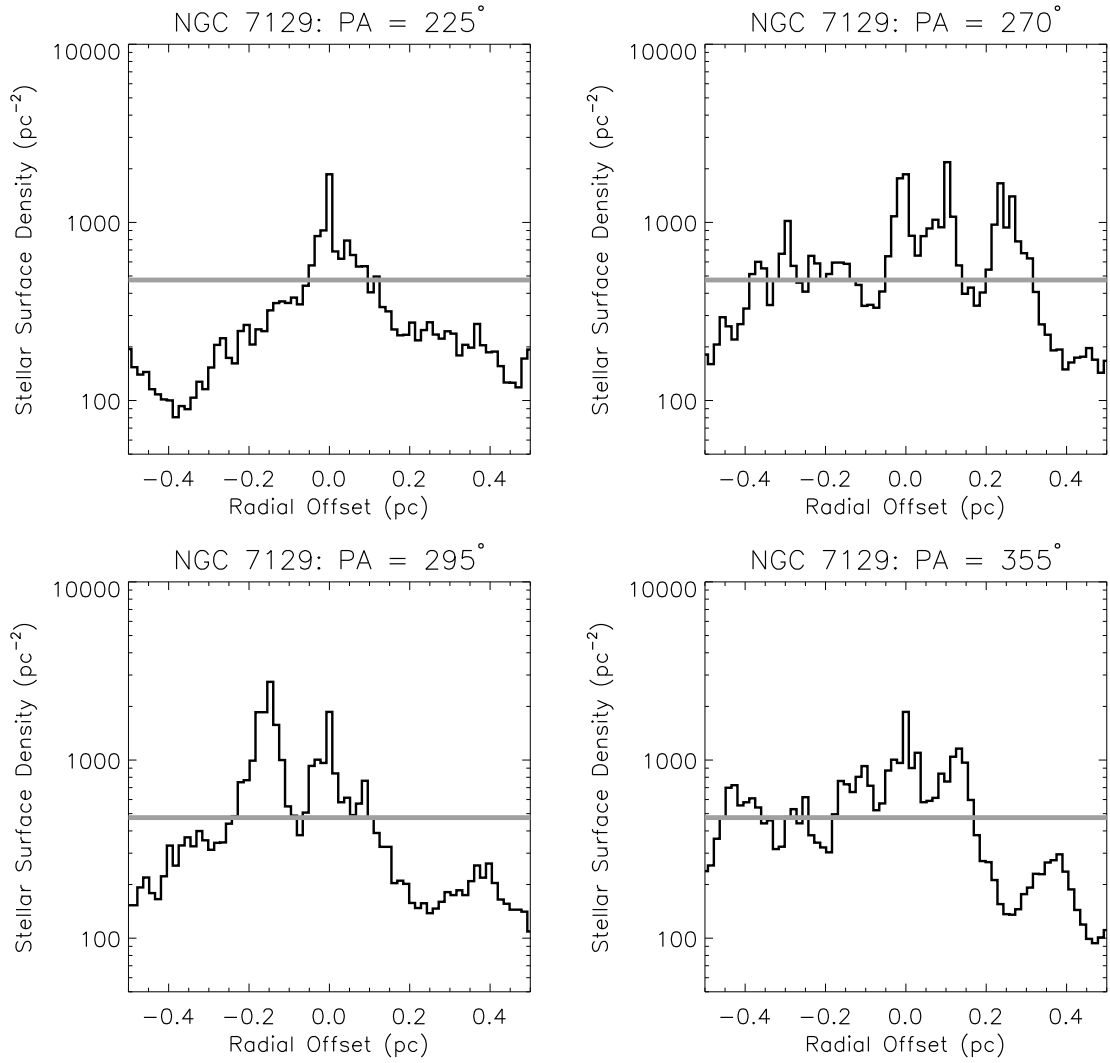}
\caption{The black lines in the above plots are stellar surface density map 
cuts for NGC~7129.  The vectors plotted in Fig.~\ref{ngc7129sg} mark 
the position, orientation, and direction of increasing offset from cluster 
center.  Note the lack of a cohesive peak and overall diffuse structure.  The 
gray line marks the $5\sigma$ above median field star density level for 
comparison, which corresponds to the first contour level in
Fig.~\ref{ngc7129sg}.\label{ngc7129slices}}
\end{figure}

The refined cluster core center point is not located at the position of peak
density in the stellar surface density map (see Fig.~\ref{ngc7129sg}), but is
instead located approximately $30^{\prime\prime}$ to the west centered on
another local density maximum.  There are three additional local density maxima
within the cluster region boundary, and all five 
have similar peak stellar surface densities (see Fig.~\ref{ngc7129slices}).  
Clearly, there is no dominant central high stellar density core in NGC~7129.  

\section{Discussion}

\subsection{The Initial Configuration and Evolution of Clusters}

As we have shown, characterizing young stellar clusters with azimuthal 
averaging methods obscures a key component in their analysis.  Those clusters 
presented that are still highly embedded, GGD~12-15 and IRAS~20050+2720, have a
high degree of asymmetry in their stellar distributions compared to NGC~7129.  
Given the similarity of the asymmetric stellar distributions to the filamentary 
structures detected in 850~$\mu$m emission, we argue that the stellar structure
observed is a direct result of asymmetry in the distribution of the natal gas.  
In contrast, the NGC~7129 cluster is approximately circularly symmetric, and 
occupies a cavity in its natal cloud.  Low average and peak stellar densities
within the cluster, 
overall lack of significant stellar density structure, and a significantly
larger cluster core radius are circumstantial evidence that recent 
expulsion of the bulk of the gas mass from the cavity has allowed the cluster 
to dynamically expand to its current state.

In the two embedded clusters there is also evidence that the star--forming gas
and dust are being disrupted.  While stellar density morphology seems related
to dust emission morphology in the embedded clusters, peak stellar densities
are often anti--correlated to dust emission maxima, suggesting that either
local extinction variation severely affects our sensitivity to stars or stellar 
feedback mechanisms are already dispersing natal material locally near high 
stellar densities.  For example, in GGD~12-15, the peak of stellar density
is located at a local minimum in 850~$\mu$m emission, suggesting that the 
central stars may be opening a cavity in the dense central 
cloud core.  Similarly, subcluster~B of IRAS~20050+2720 is just outside the 
region of peak 850~$\mu$m emission, possibly indicating that it too has 
dispersed much of its gas, as suggested by \citet{chen97}.  Subcluster~B may be 
entering a phase of dynamical expansion similar to what we suggest to explain
the current state of the NGC~7129 cluster, while the remaining filament
is still actively forming stars in subclusters~A~and~C and the small millimeter
emission cores detected nearby \citep{chin01}.

Analytical investigations and numerical simulations of the dynamics of young
clusters following rapid expulsion of their natal gas have shown that dynamical
expansion and dissolution can occur on short time scales in young clusters
\citep{am01,kpm99}.  Indeed, over the course of 1~Myr an unbound star moving
at 1~km/s can travel 1~pc, suggesting that the
high stellar densities observed in young clusters can only last while the
system is gravitationally bound by the mass of the natal molecular cloud.  
The observations presented in this work further these arguments.  Understanding
the impact of forming stars on their parent cloud is crucial to understanding
the complex dynamics of 
embedded cluster evolution.  By analyzing stellar density distribution 
morphology in relation to molecular cloud structure, observational analyses can
more adequately address the clear link between star formation, gas expulsion,
and the dynamics of the clusters and how these processes guide the evolution of young clusters.

\subsection{The Impact of Clustering on Disk and Envelope Evolution}

Simulations of massive protostellar disk interactions suggest that truncation
and fragmentation of the disks are likely for close approaches with other stars
on the order of the disk radius \citep[e.g.,][]{boff98}.  A recent
investigation into the effects of close approaches on classical T~Tauri
disks as a function of a variety of parameters suggests tidal truncation and
mass loss are possible, as well as enhanced accretion \citep{pfal05}.
Furthermore, the gravitational instability planet formation scenario may
require a close approach with another star to perturb the protoplanetary disk
to begin the process \citep{boss02}.  
If close approaches are likely for a significant fraction of pre-main sequence
stars in an embedded cluster, this would have a profound effect on disk
evolution and perhaps planet formation in a clustered star--forming
environment.  

Observations other than those presented here provide further clues that
envelope and disk interactions may happen frequently and indeed may be
significant in the evolution of disks and the formation of planets.  Several
disks around pre-main sequence stars in the Orion Nebula Cluster (ONC) have
been directly
observed with {\it HST} \citep{mo96}; these disks have radii ranging from 
as small as 50~AU to as large as 1000~AU, and their central stars have
derived ages of 0.8-3.0~Myr.  Given the high stellar densities measured in the
ONC, as well as the wide range of disk radii and sharp disk edges observed,
tidal truncations from close encounters are a likely explanation 
\citep{mo96}.  A 10~AU gap in the disk of $\sim$1~Myr old T~Tauri star
CoKu Tau/4 has been discovered in observations from the {\it Spitzer} 
Infrared Spectrograph, suggesting the presence of a planet
\citep{forr04,dale04}.  This suggests that giant planet formation may occur
very early in a protoplanetary disk's lifetime, an argument in favor of planet
formation via gravitational instability.  
Current orbital models for the
recently discovered distant minor planet Sedna \citep{btr04} argue for a close
interaction from a passing star, suggesting that our solar system may have
formed in a clustered environment.  

\begin{figure}
\epsscale{1}
\plotone{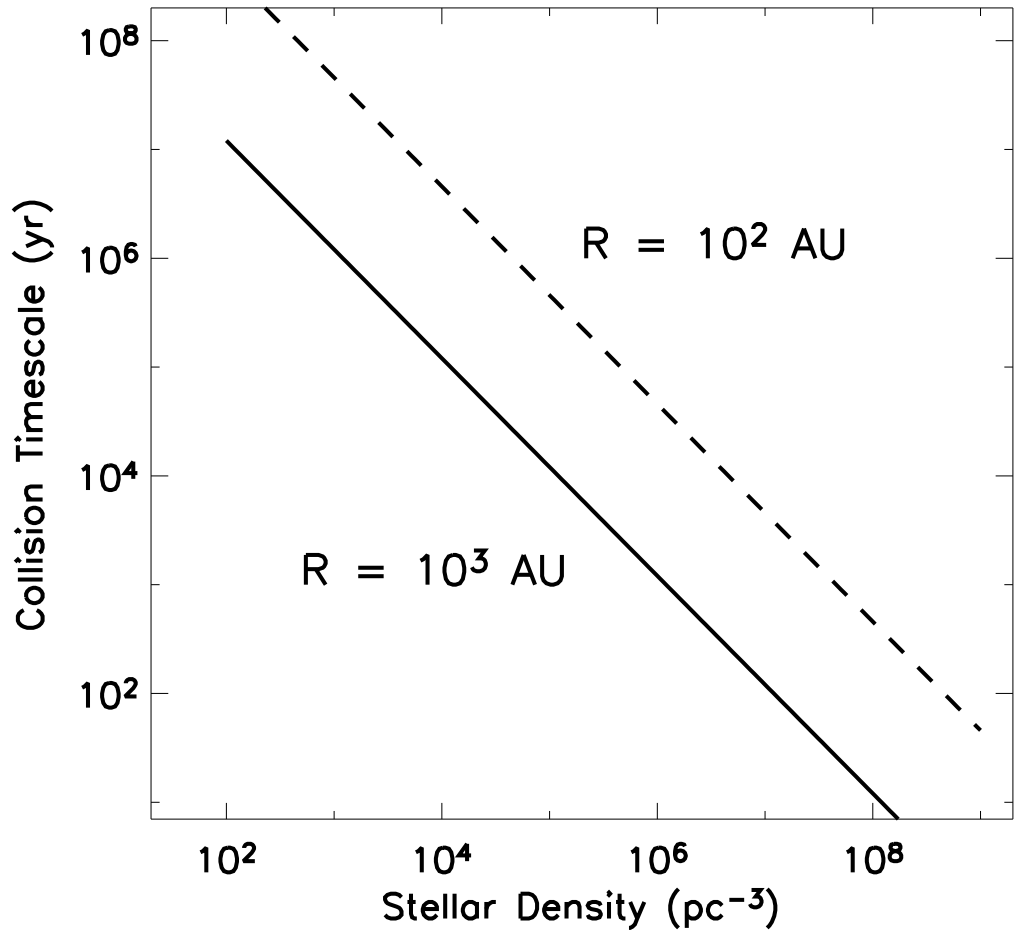}
\caption{Estimates for collisional timescales for protostellar envelopes
($R = 10^3$~AU, solid line) and T~Tauri disks ($R = 10^2$~AU, dashed line) as a 
function of stellar volume density.\label{colltime}}  
\end{figure}

One of the strengths of the stellar surface density mapping method adopted for
this work is its ability to probe stellar densities over a wide dynamic
range with adequate spatial resolution.  While stellar collisions are very 
unlikely at the stellar densities reported in this work 
\citep[cf.][and note that lower average stellar masses further increase the
timescales in Figure~1 of that study]{bb02}, close approaches on the order of
protostellar envelope or large T~Tauri disk radii
\citep[$R = 10^3$~AU,][]{ma01,mo96} at these densities
may be much more common.  The time--scale for a given star to
experience a close encounter when among a cluster of stars of stellar density
$n$ and velocity dispersion $v_{disp}$ can be roughly approximated via a simple
analysis outlined in \citet{bt87}.  They express this time--scale, $t_{coll}$,
as:
\begin{displaymath}
\frac{1}{t_{coll}} = 16\sqrt{\pi}nv_{disp}R_*^2(1+\frac{GM_*}{2v_{disp}^2R_*})
\end{displaymath}
All stars are assumed to have the same mass, $M_*$, and the same interaction 
radius, $R_*$.  For this analysis, we assume $M_* = 0.5~M_{\odot}$ as an
appropriate median stellar mass from young cluster initial mass functions
reported in the literature \citep[e.g., ][]{muen02,luhm03}, and
$v_{disp} = 1~km~s^{-1}$ based
on Maxwellian velocity dispersions derived from C$^{18}$O linewidths for these
clusters (see Table~\ref{littable}).  Figure~\ref{colltime} shows the derived
collisional timescales as a function of stellar density for two choices of 
$R_*$, $10^3$~AU for protostellar envelopes \citep{ma01} and large T~Tauri
disks and $10^2$~AU for classical T~Tauri disks \citep{mo96}.  The timescales
considered must be less than or on the order of $10^5$ to $10^6$~yr to allow
for a significant frequency of interactions to occur over the length of a
forming star's protostellar collapse phase \citep{kh95} or the approximate
length of a young cluster's active star--forming phase \citep{ps00},
respectively.  For example, at a density of $10^4$~pc$^{-3}$, protostars and
T~Tauri stars with large disks are likely to have a close encounter on the
order of their envelope or disk radius on a timescale of $\sim$$10^5$~yr, while
classical T~Tauri disk radius interactions occur much less often, on a 
timescale of $\sim$$10^7$~yr.  The latter timescale suggests that $\sim$10$\%$
of stars at densities of $10^4$~pc$^{-3}$ would have a close encounter on the
order of $10^2$~AU in $\sim$1~Myr, if the high stellar density environment can
indeed last that long.  

Utilizing our stellar density maps and cluster membership and field star  
contamination analyses, we can estimate the prevailing stellar density 
environment around the stars in the cluster regions relative to the interaction 
criteria derived above.  First, the associated stellar volume density estimate
is determined for each star within the cluster region boundary using the
nearest measurement in our density maps.  Note that these are nearest neighbor
derived volume density estimates, thus they may be overestimated.  
We count the number of stars in the cluster regions with densities greater than 
$10^4$~pc$^{-3}$.  The mean number of field stars falling within the high
density areas of the cluster region is estimated directly from the
$N_{CM}$ field star models.  
See Table~\ref{densitytable} for results for each cluster.

In the two highly embedded clusters presented, GGD~12-15 and IRAS~20050+2720, 
$72\%$ and $91\%$ of the inferred cluster region members respectively are in
locations with stellar densities above $10^4$~pc$^{-3}$.  
This result suggests that there is ample opportunity for protostellar envelope
and large T~Tauri disk interactions to occur in these two clusters even over
relatively short timescales ($<10^5$~yr).  Furthermore, while classical T~Tauri
disk scale interactions are less common, they still may occur in as many as
$10\%$ of the stars in these two clusters over the course of $\sim$1~Myr.
Indeed, 1~Myr is probably an upper limit on the high stellar density phase
lifetime of these clusters, as significant gas expulsion is already apparent.
Regardless, this estimate is consistent with
the suggestion of \citet{acp03} that up to $30\%$ of T~Tauri disks must be
dispersed within 1~Myr, possibly due to a combination of close encounters with
other stars and more gradual evolution and accretion in the smallest disks.  
In contrast, less than $24\%$ of the stars in the cluster core boundary of
NGC~7129 are at densities greater than $10^4$~pc$^{-3}$.  Clearly, there is
little chance for further disk-affecting interactions in this cluster, although
this is not entirely unexpected.  \citet{gute04} report the disk fraction in
the cluster region is $54\%$, suggesting that the population is already
somewhat evolved.  Furthermore, spectroscopic evidence suggests that the low
mass stars in NGC~7129 are between 1.5-2.0~Myr old \citep{hill95}.  

If we assume cluster asymmetry and high average extinction are signs of extreme
youth ($<1$~Myr) in the young clusters presented, then the high density
environments we observe in GGD~12-15 and IRAS~20050+2720 provide evidence
supporting the idea that protostellar envelope and large ($10^3$~AU) T~Tauri
disks are likely to be affected by close encounters with other members in the
youngest clusters, and this may lead to significant tidal stripping or
conversely, accelerated accretion, early in their evolution.  If the molecular
gas mass of the natal cloud is indeed adequate to maintain the high density
configurations we have observed over a 1~Myr timescale, a small but measurable
fraction of the classical T~Tauri disks have the chance to
be significantly truncated or disrupted entirely.  Once gas expulsion and
subsequent dynamical expansion occurs, as we argue has happened in NGC~7129, 
stellar densities decrease to the level at which the disks are allowed 
to evolve and eventually disperse by means of more gradual activity, such as 
photoevaporation \citep{acp03}, residual accretion, or planet formation.  
Clearly, determining the timescale of bulk gas expulsion is of fundamental
importance to understanding the broad implications disk-affecting close
encounters may have for disk fraction studies, planet formation frequency and
timescale, and the stellar mass spectrum.

\acknowledgments

FLAMINGOS was designed and constructed by the IR instrumentation group
(PI: R. Elston) at the University of Florida, Department of Astronomy
with support from NSF grant (AST97-31180) and Kitt Peak National Observatory.
This research has made use of the NASA/IPAC Infrared Science Archive, which is
operated by the Jet Propulsion Laboratory, California Institute of Technology,
under contract with the National Aeronautics and Space Administration.
This publication makes use of data products from the Two Micron All Sky 
Survey, which is a joint project of the University of Massachusetts and 
the Infrared Processing and Analysis Center/California Institute of 
Technology, funded by the National Aeronautics and Space Administration 
and the National Science Foundation.
This research has made use of the SIMBAD database, operated at CDS, 
Strasbourg, France.  JPW acknowledges support from NSF grant AST-0324323.

\clearpage

\begin{deluxetable}{rccc}
\tablecaption{Cluster characteristics compiled from the literature.\label{littable}}
\tablehead{
\colhead{ } & \colhead{GGD~12-15} & \colhead{IRAS~20050+2720} & \colhead{NGC~7129}
}
\startdata
Cluster Distance (pc): & 830\tablenotemark{a} & 700\tablenotemark{b} & 1000\tablenotemark{c} \\
Molecular Cloud Mass\tablenotemark{d} ($M_{\odot}$): & 745 & 275 & 400 \\
Velocity Dispersion\tablenotemark{d} ($km~s^{-1}$): & 0.92\tablenotemark{e} & 1.07\tablenotemark{e} & 0.62\tablenotemark{e} \\
Far IR Luminosity\tablenotemark{d} ($L_{\odot}$): & 5680 & 227 & 1680\tablenotemark{f} \\
\enddata
\tablenotetext{a}{\citet{hr76}}
\tablenotetext{b}{\citet{wilk89}}
\tablenotetext{c}{\citet{raci68}}
\tablenotetext{d}{\citet{ridg03}}
\tablenotetext{e}{Converted from FWHM values ($\sigma=FWHM/2.35$) reported in \citet{ridg03}.}
\tablenotetext{f}{Adjusted to account for different adopted distance.}
\end{deluxetable}

\begin{deluxetable}{rccc}
\tablecaption{Point Source Detection Completeness Limits (90\%)\label{complete}}
\tablehead{
\colhead{ } & \colhead{GGD~12-15} & \colhead{IRAS~20050+2720} & \colhead{NGC~7129}
}
\startdata
$K$: & 18.25 & 17.20\tablenotemark{a} & 17.90\tablenotemark{a} \\
$H$: & 18.85 & 18.20 & 18.70 \\
$J$: & 20.00 & 18.70 & 19.25 \\
\enddata
\tablenotetext{a}{$K_S$}
\end{deluxetable}

\begin{deluxetable}{rccc}
\tablecaption{Cluster core characteristics.\label{clustertable}}
\tablehead{
\colhead{ } & \colhead{GGD~12-15} & \colhead{IRAS~20050+2720} & \colhead{NGC~7129}
}
\startdata
Cluster-Dominated Region Radius\tablenotemark{a} (pc): & 0.29 & 0.29 & 0.59 \\
Cluster Core Radius\tablenotemark{a} (pc): & 0.24 & 0.24 & 0.65 \\
No. of Cluster Region Members: & $98\pm10$ & $91\pm10$ & $122\pm16$ \\
No. of Cluster Region Field Stars: & $15\pm2$ & $58\pm4$ & $114\pm8$ \\
Azimuthal Asymmetry Parameter: & 1.588 & 1.421 & 1.207 \\
Peak Surface Density\tablenotemark{b} (pc$^{-2}$): & 5910 & 6320 & 2750 \\
Mean Surface Density\tablenotemark{c} (pc$^{-2}$): & 371 & 344 & 91.9 \\
Peak Volume Density\tablenotemark{b} (pc$^{-3}$): & $2.7 \times 10^5$ & $3.0 \times 10^5$ & $8.6 \times 10^4$ \\
Mean Volume Density\tablenotemark{c} (pc$^{-3}$): & 959 & 891 & 106 \\
\enddata
\tablenotetext{a}{The cluster-dominated region radius, often more simply called "cluster radius" in the text, is used throughout this work in lieu of the cluster core radius.} 
\tablenotetext{b}{Peak densities are derived from nearest neighbor distances (see Section~\ref{gather}).}
\tablenotetext{c}{Mean densities are derived from estimated cluster memberships and cluster-dominated region radii.}
\end{deluxetable}

\clearpage

\clearpage

\clearpage

\clearpage

\clearpage

\clearpage

\begin{deluxetable}{rccc}
\tablecaption{Nearest Neighbor Volume Density Environments by Cluster\label{densitytable}}
\tablehead{
\colhead{ } & \colhead{GGD~12-15} & \colhead{IRAS~20050+2720} & \colhead{NGC~7129}
}
\startdata
No. of Stars at $>10^4$~pc$^{-3}$: & 74 & 100 & 35 \\
Est. No. of Field Stars: & 3 & 17 & 6 \\
No. of Members at $>10^4$~pc$^{-3}$: & 71 & 83 & 29 \\
Fraction of total in core: & 72.4\% & 91.2\% & 23.8\% \\
\enddata
\end{deluxetable}

\end{document}